\documentclass[10pt, sigconf]{acmart}
\usepackage{booktabs} % For formal tables
\usepackage{mathtools}
\usepackage{bm}
\usepackage{algorithmic}
\usepackage{graphicx}
\usepackage{textcomp}
\usepackage{xcolor}
\usepackage{subcaption}
\usepackage{diagbox}
\usepackage{tabulary}
\usepackage{mathtools}

\setcopyright{rightsretained}
\acmDOI{}

\acmISBN{}

\acmConference[]{}{}{}
\copyrightyear{2020}

\begin{document}
%-------------------------------------------------------------------------------

%don't want date printed
\date{}

\title{\Large \bf A Multi-Armed Bandit-based Approach \\
to Mobile Network Provider Selection}

%for single author (just remove % characters)
%\author{
%{\rm Your N.\ Here}\\
%Your Institution
%\and
%{\rm Second Name}\\
%Second Institution
% copy the following lines to add more authors
% \and
% {\rm Name}\\
%Name Institution
%} % end author

\author{ \small Thomas Sandholm and Sayandev Mukherjee } 
\affiliation{
  \institution{Next-Gen Systems, CableLabs}
}

% !TEX root = aim.tex

\begin{abstract}
We argue for giving users the ability to ``lease'' bandwidth temporarily from any mobile network operator.  We propose, prototype, and evaluate a spectrum market for mobile network access, where multiple network operators offer blocks of bandwidth at specified prices for short-term leases to users, with autonomous agents on user devices making purchase decisions by trading off price, performance, and budget constraints.

We begin by showing that the problem of provider selection can be formulated as a so-called Bandit problem.  For the case where providers change prices synchronously, we approach the problem through contextual multi-armed bandits and Reinforcement Learning methods like Q-learning either applied directly to the bandit maximization problem or indirectly to approximate the Gittins indices that are known to yield the optimal provider selection policy.  We developed a simulation suite based on the open-source PyLTEs Python library.  For a simulated scenario corresponding to a practical use case, our agent shows a $20-41$\% QoE improvement over random provider selection under various demand, price and mobility conditions.  Finally, we show that the problem of provider selection for a given user agent in the general spectrum market with asynchronously changing prices can be mathematically modeled as a so-called dual-speed restless bandit problem.  

We implemented a prototype spectrum market and deployed it on a testbed, using a blockchain to implement the ledger where bandwidth purchase transactions are recorded. User agents switch between provider networks by enabling the corresponding pre-downloaded eSIM profile on the devices on which these agents live.  The real-life performance under different pricing, demand, competing agent and training scenarios are experimentally evaluated on the testbed, using commercially available phones and standard LTE networks. The experiments showed that we can learn both user behavior and network performance efficiently, and recorded $25-74\%$ improvements in QoE under various competing agent scenarios.
\end{abstract}

\maketitle

% !TEX root = aim.tex

\section{Introduction}\label{sec:introduction}
\subsection{The operator landscape today: MNOs and MVNOs}
Today mobile network provisioning and operation with bandwidth guarantees 
requires 1) a license to operate on a dedicated band frequency range,
2) permission to install radio transceivers in strategic locations
and 3) infrastructure to connect the transceivers to a core network backhaul.
Each of these requirements can be used to differentiate a service, but
at the same time also serves as a roadblock for providing new services.    
The end-result is inefficient (both in terms of utilization, performance and cost) 
use of network resources, such as RF spectrum across different locations and time periods.  
The most common way to address these issues today is through peering and roaming 
agreements between primary operators or spectrum license holders, a.k.a. Mobile
Network Operator (MNOs), and secondary providers, network resource lessees, a.k.a Mobile
Virtual Network Operators (MVNOs). Traditionally, these arrangements were set up to
improve coverage of a service. More recently, a new type of MVNO has emerged
that allows operation on multiple MNOs' networks in a given location to improve
performance as well as coverage, e.g. GoogleFi.

From an end-user perspective these new MVNOs operate similarly to services offered from
a traditional MNO or MVNO. Contracts follow the traditional monthly or yearly agreements,
and the user controls neither the network used at any given time nor the set of networks
that can be selected from at any given time and location.  More importantly, an aggregator-MVNO's (such as GoogleFi's) decision as to which network is best at a particular time and place is based on an aggregate utility over all its served users, and does not take the budget or willingness-to-pay preferences of any individual user into account
given a task at hand.

\subsection{eSIMs and the promise of agency to end-users}
With the introduction of eSIMs, end-users can pick and choose from a large number of competing
network providers in any given location without having to physically visit a store or wait for
a physical SIM card to be shipped. Providers of eSIMs typically offer shorter contracts with limited 
data volumes. Modern phones allow both a physical and eSIM to be installed side-by-side,
and any number of eSIM profiles may be installed on the same device, albeit
currently only a single one can be active at any given time. This setup is ideal for travelers
or for devices that just need very limited connectivity over a short period of time.
An eSIM profile may also be used to \emph{top off} primary plans and avoid hitting bandwidth
throttling thresholds.

Currently, end-users have to manually switch between SIMs and eSIMs to activate, although 
different profiles can be designated for phone calls and messaging and data for instance.

\subsection{The provider selection problem}
One could argue that simply switching to the provider with the best signal at any given
time is sufficient, but apart from switching cost it could also be suboptimal because
estimating future signal strength in a complex mobile network is non-trivial and
the most recent signal measured is not necessarily the best predictor~\cite{herath2019}.
Mobility also plays an important role as the Quality of Service (QoS) offered depends heavily on location, and therefore
machine learning models have been deployed to improve performance in mobile networks
by predicting the best base stations to serve a user given a mobility pattern~\cite{wick2017}.
Different applications may have different networking needs, complicating the selection process further,
and motivating work such as Radio Access Technology (RAT) selection based on price and capacity~\cite{passas2019}.  Finally, the user may also impose constraints on the budget available for bandwidth purchases, both in the near-term (daily or weekly) and the longer-term (monthly).

\subsection{A learning agent approach to provider selection}
In this work we focus on data usage and allowing an agent on the device to determine which network
provider (e.g. eSIM) to use at any given time and location to optimize both bandwidth delivered and cost
incurred. Because the agent is deployed on the device, it has access to the device location and
the task being performed (i.e. the bandwidth demand of the current app) in addition to the traditional network signal strength measures used to
pick a provider. Furthermore, since our agent is local this privacy sensitive data never has to
leave the device, and can be highly personalized to the behavior of the user over a long 
period of time. 

Instead of relying on QoS guarantees offered by service providers and outlined in fine-print in obscure legal contracts, the agent learns the \emph{Quality of Experience} (QoE), here defined as bandwidth per unit price with potential upper and lower bounds, for each provider under different demand conditions by exploration. The exploration itself follows a \emph{learning} approach to construct an optimal switching policy between providers given a certain state, activity and budget of the user.
We evaluate our solution both via simulation and with commercial smart phones on an experimental testbed.

\subsection{MDPs, Bandit problems, and Reinforcement Learning}
We will show in Sec.~\ref{sec:reward_principle} that the problem of how the user agent learns the best tradeoff between the exploration of different provider selection policies versus exploiting provider selections that have worked well in the past may be posed as a so-called \emph{Bandit problem}~\cite{lattimore2020}. Bandit problems are a widely-studied class of \emph{Markov Decision Process} (MDP) problems. If the dynamics of the interactions between the agent and the environment (the networks and other user agents), as represented by various probability distributions, are fully known, then the provider selection problem corresponds to a Bandit problem that can be solved exactly. However, this is seldom the case.  A different discipline for attacking MDP problems when the probability distributions are unknown, called \emph{Reinforcement Learning} (RL)~\cite{sutton2018}, can be applied to obtain solutions to these Bandit problems~\cite{duff1995}, often using an algorithm called \emph{Q-learning}~\cite[Sec.~6.5]{sutton2018}. We will discuss and illustrate these approaches to the provider selection problem for a practically useful scenario in Sec.~\ref{sec:level2}.

\subsection{Outline of the paper}
In Sec.~\ref{sec:relatedwork} we provide a survey of related approaches in the literature, compiled across multiple fields, and compare and contrast the approach in the present work against that of other authors. Sec.~\ref{sec:motivation} describes an experimental study that clearly illustrates the benefits of provider selection, assuming of course that such provider selection is supported by a spectrum market in short-term leases for access bandwidth sold by various network operators.  

In Sec.~\ref{sec:market} we describe an abstract version of such a spectrum market in a way that lets us mathematically formulate the provider selection problem.  In Sec.~\ref{sec:provider_selection} we show that this problem formulation is an example of a Bandit problem.  In Sec.~\ref{sec:mab_review}, we provide a brief review of approaches to solve the Multi-Armed Bandit problem.

In Sec.~\ref{sec:level2}, we theoretically formulate the version of the Multi-Armed Bandit problem wherein the agent selects between multiple fixed-price plans where the assigned bandwidth to a user is dependent on the number of other users who select the same provider (not disclosed to the user).  Sec.~\ref{sec:simulation} describes a simulation setup to evaluate the algorithms proposed in Sec.~\ref{sec:level2}.

In Sec.~\ref{sec:implementation}, we delve into the details of the design and implementation of the spectrum market that we previously defined abstractly in Sec.~\ref{sec:market_defn}.  This spectrum market is deployed on a testbed, on which we then performed experiments to evaluate the performance of the algorithms proposed in Sec.~\ref{sec:level2}, and compare them against the simulation results we obtained in Sec.~\ref{sec:simulation}.  Sec.~\ref{sec:experiment} describes the experimental setup and discusses the experimental results.

We describe the mathematical model for the general provider selection problem in the general spectrum market in Sec.~\ref{sec:dualspeedrestless}.  Finally, we summarize our findings and conclusions in Sec.~\ref{sec:conclusions}.

% !TEX root = aim.tex

\section{Related Work}\label{sec:relatedwork}

Optimal service provider selection has been investigated in several domains,
such as Cloud computing~\cite{wei2019, barrett2013}, telecommunications~\cite{haddar2018, trestian2012}, 
wireless infrastructure service providers~\cite{vamvakas2018, salem2005}, 
HetNets~\cite{amine2018, tan2014} and WLANs~\cite{bojovic2011, sandholm2018}.
There is also a large body of work on provider-focused centralized optimization 
of performance of users connected to one or many base stations using cognitive
radio technology~\cite{mismar2019, mosleh2020, nader2019, zhang2018}.

\subsubsection*{Cloud computing}
In ~\cite{wei2019, barrett2013} the authors address the problem of selecting an optimal Cloud infrastructure
provider to run services on using Q-learning techniques. 
In~\cite{wei2019}, the authors define an RL reward function where they consider the profit
from offering the service as well as the idle time of virtual machines. They consider whether
to purchase reserved instances or on-demand instances with different pricing schemes with the goal
of auto-scaling based on a stochastic load setting. Although their high-level RL mechanism is similar to ours,
our work differs in multiple aspects beyond the application domain. We focus on throughput optimization,
we assume a sunk cost for a time expiring capacity purchase and consider not only the workload but the QoS offered as well to be stochastic.

\subsubsection*{Telecommunications}
The problem of selecting the best telecom service provider for VOIP service is investigated in ~\cite{haddar2018}.
They take QoS, availability and price into account using a decision-tree model designed to predict 
classes of service levels most appropriate for users given their requirements. The method is a dynamic rule-based
solution and hence relies on finding good delineating features, and having a large supervised training data set.
In contrast, our approach can learn over time, and adjust more easily to non-stationary behavior. Our approach does require training, but not supervised or manual classification of samples. In ~\cite{trestian2012} a wide range of game-theoretical approaches to mobile network provider selection are surveyed, with the conclusion that computational complexity
may hamper its adoption in practice. Moreover, claiming Pareto optimality may be difficult in an
non-cooperative environment where the pricing mechanism is not centrally controlled.  

\subsubsection*{Wireless infrastructure service providers}
Similar to our work,~\cite{vamvakas2018} also considers mobile customer wireless service selection, albeit focusing on Wireless Internet Service Providers (WISPs) as opposed to eSIM providers.
Their primary focus is on power allocation as opposed to bandwidth optimization. Moreover, they rely on
global state to be communicated about each user and provider to find a global optimum (Nash Equilibrium).
In contrast, our RL algorithm only needs local information and learns which providers to pick from experienced QoS.
The authors in~\cite{vamvakas2018} propose a learning automata model that seeks to model switching probabilities
instead of prescribing switch or stay actions, as in our approach.
A reputation-based trust service and new security protocols were proposed to solve the 
problem of WISP selection in~\cite{salem2005}.
Adding a new trusted component to the network is, however, a steep hurdle to adoption. On the other hand, a poorly performing
provider in our scenario would get less likely to be picked in the future due to deteriorating consumer-perceived
rewards fed into the reward function in our approach.

\subsubsection*{HetNets}
A wireless service provider may decide to serve an area by offering an LTE macro cell and a mix of individual
smaller contained LTE pico cell base stations and Wi-Fi access points (APs), in what is typically referred to as a Heterogeneous Network or HetNet, for short. 
Each mobile station can then pick which base station (BS) as well as which technology to use based on bandwidth requirements,
capacity available, and SINR. The problem of picking the best AP or BS for a user investigated in~\cite{amine2018}
is similar to
our problem of network service provider selection, although~\cite{amine2018} generally assumes that more information
is available about competing users that can be used in centralized decisions. The authors of~\cite{amine2018} propose a genetic algorithm
to solve a multi objective optimization problem and show that it does significantly better than methods simply
based on maximizing the SINR. Our solution, on the other hand, is fully decentralized and applies an exploration-exploitation
process to determine which provider offers the best QoS over time. In~\cite{tan2014} the authors propose
a Q-learning algorithm to allow users to select optimal small cells to connect to without the need of a central
controller. Their utility or reward function does not, however, take local conditions into account, 
such as the requirements of the currently running application. Furthermore, they don’t consider price and thus not QoE, 
assume a single provider, use a Q-learning instead of bandit algorithm (we will show below that the latter 
outperforms the former). 

\subsubsection*{WLANs}
In a Wireless Local Area Network (WLAN) context the problem can be defined as a mobile 
station selecting the best AP that is in signal range based on past experience. 
In~\cite{bojovic2011} a multi-layer feed-forward neural network model is proposed to
learn to predict the best provider for an STA given various WLAN parameters such as
signal to noise ratio, failure probability, beacon delay, and detected interfering
stations. In contrast to our approach, the neural network in~\cite{bojovic2011} is trained via supervised learning model, relying on a large set of labeled training data. Moreover,
it does not take cost into account and assumes all APs provide the same service
given the detected signal inputs. In ~\cite{sandholm2018} an approach
to user association is proposed that learns the optimal user-to-AP mapping
based on observed workloads. It differs from the approach presented in the present work
in that it relies on a central controller. 

\subsubsection*{Cognitive Radio Networks}
In addition to service provider selection, there has also been a lot of research into cognitive network
operation and self-configuring APs or BSs to solve resource allocation problems centrally. 
Many of these algorithms are based on RL and Q-learning techniques, e.g. ~\cite{mismar2019, mosleh2020, zhang2018}.
RL and related Multi-armed bandwidth techniques have also been deployed to do link scheduling, 
determining whether to be silent or transmit in a time slot,
across a distributed set of independent transmitters~\cite{nader2019, kang2018}.

\subsubsection*{Dynamic channel selection}
The general mathematical formulation of the dynamic channel selection problem as a Markov Decision Process yields a Restless Multi-armed Bandit Problem (RMBP).  Unfortunately, there is no closed-form solution to this problem aside from special cases~\cite{liu2010, duran2018}.  Q-learning techniques have been proposed in theory~\cite{avrachenkov2020} and implemented via deep learning models~\cite{wang2018, iclr2021}, but the resulting model complexity, both in computation and storage requirements, is too large to be suitable for deployment as a user agent on a mobile device.  The closest approach to ours is in~\cite{wang2018}, but in contrast to their choice of a very simple reward function and a sophisticated deep-learning approximation to compute the action-value function, we use a more sophisticated reward function but a relatively simple action-value function that can be implemented as a table.

% !TEX root = aim.tex

\section{Motivation}\label{sec:motivation}
Our hypothesis that there is an opportunity
to optimize network selection relies on different network providers
offering different levels of QoS in different locations
at different times in non-trivial ways. To measure QoS
differences in different locations, we measured the 
performance of two tasks for two different mobile
network providers in 40 locations across four
neighboring cities. 

The first task is a live video-conferencing session and
represents a real-time interactive use case. The second
task is a photo upload and represents a throughput bound
batch use case. For the first task we measure frames
decoded, frames dropped and connection time. For the
second task we measure time to upload a photo to
a Cloud server. For each location we also measure the
signal strength with each network provider.

Table~\ref{T:motivation_summary} summarizes the differences
between sticking to a single provider versus dynamically picking 
the best provider across the metrics measured (for each task and location).

\begin{table}[htbp]
        \caption{Difference between best provider and fixed provider. Improvement opportunity in bold.}
\begin{center}
\begin{tabulary}{\linewidth}{|L|L|L|}
\hline
	\textbf{Metric} & \textbf{Provider} & \textbf{Difference (\%)}\\
\hline
Decoded Frames & Ubigi & $5.33$  \\
	& GigSky & $\mathbf{0.33}$  \\
\hline
Dropped Frames & Ubigi & $\mathbf{-41.20}$  \\
 & GigSky & $-57.68$  \\
\hline
Connection Time & Ubigi & $-10.54$  \\
	& GigSky & $\mathbf{-0.23}$  \\
\hline
Upload Time & Ubigi & $-40.36$  \\
& GigSky & $\mathbf{-26.60}$  \\
\hline
Signal Strength & Ubigi & $\mathbf{11.20}$  \\
 & GigSky & $20.87$  \\
\hline
\end{tabulary}
\label{T:motivation_summary}
\end{center}
\end{table}

The differences are measured as $\frac{b-p}{p}$ where $b$
is the measurement for the best provider and $p$ is the
measurement for the fixed provider. We also mark the
improvement over the best fixed provider in bold as an indicator
of the opportunity for picking the best provider in 
a location. We note that the {\it Dropped Frames}, 41+\%, from
the video-conference task and the {\it Upload Time}, 26+\%,  metric
from the photo upload task show the greatest opportunities
for improvement. The signal strength opportunity is also
significant (11+\%).

\begin{figure}[htbp]
        \centerline{\includegraphics[scale=0.1]{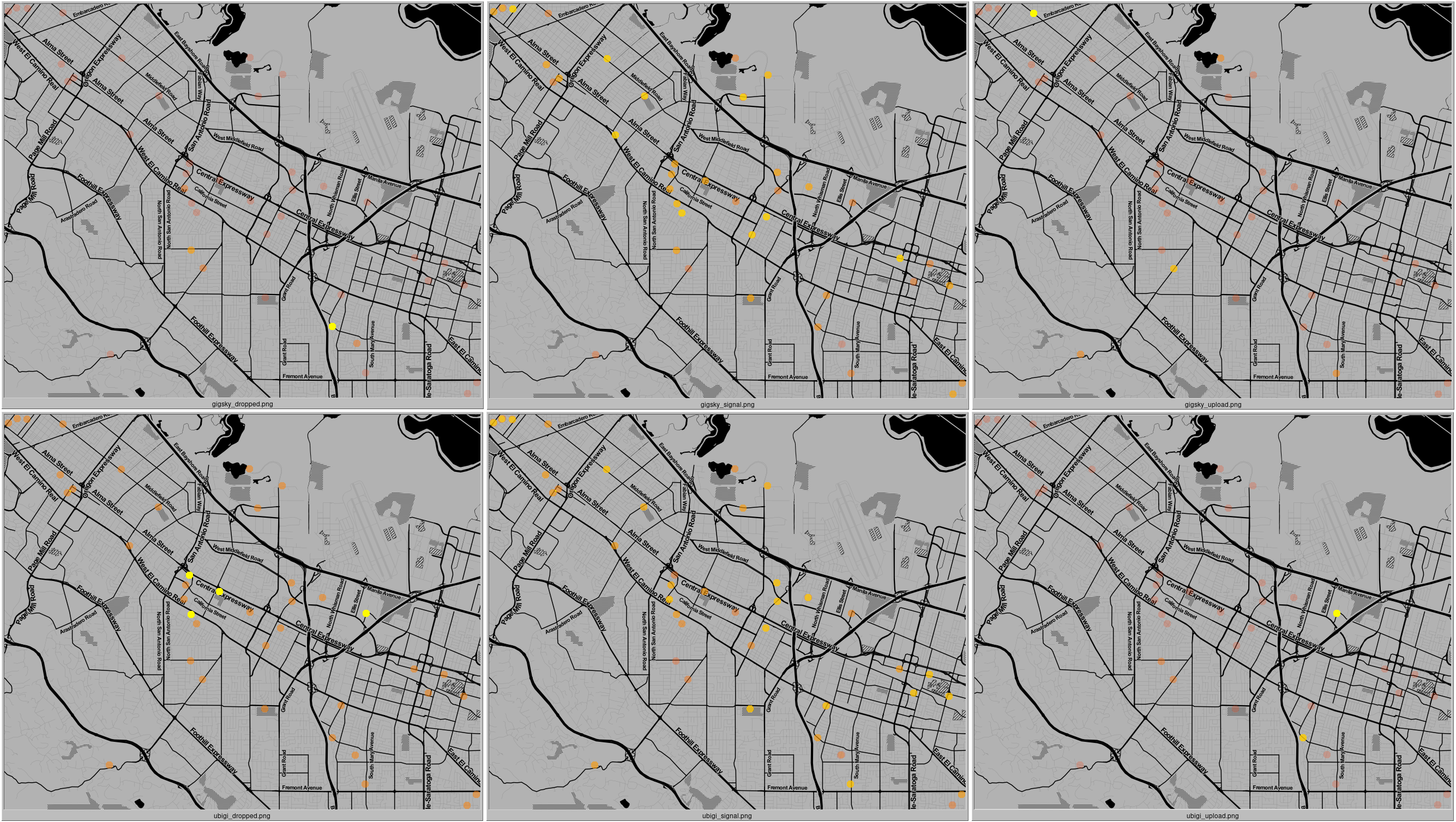}}
	\caption{Gigsky Provider Measurements (top row) and Ubigi Provider Measurements (bottom row).
	Metrics from left to right, dropped frames, signal strength, and upload time.}
\label{heatmap}
\end{figure}

To visualize the distribution of QoS levels across locations for
different providers we mark each location with a heat point,
where a higher value quantized on a scale from 1-5 gets a more
yellow point. Figure~\ref{heatmap} exemplifies a non-trivial mapping
of the best provider in each location.

Now, to quantify the relationships between different metrics we compute
the correlations between each metric and provider. The correlations
could, for example, give a hint whether a higher
frame drop rate could be explained by a poorer signal, or whether the location
could give a hint has to which provider would have the best signal.

\begin{figure}[htbp]
        \centerline{\includegraphics[scale=0.5]{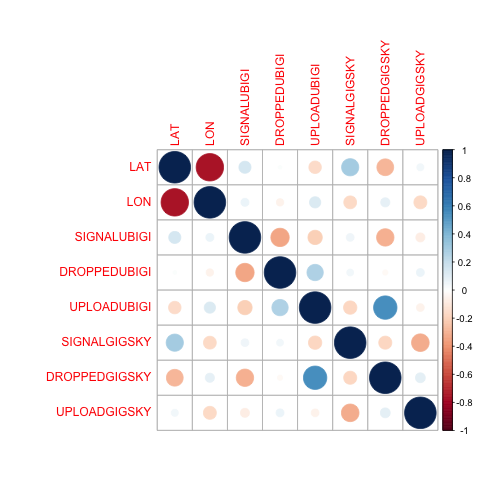}}
	\caption{Correlations between metrics and providers.}
\label{correlation}
\end{figure}

\begin{figure}[htbp]
        \centerline{\includegraphics[scale=0.5]{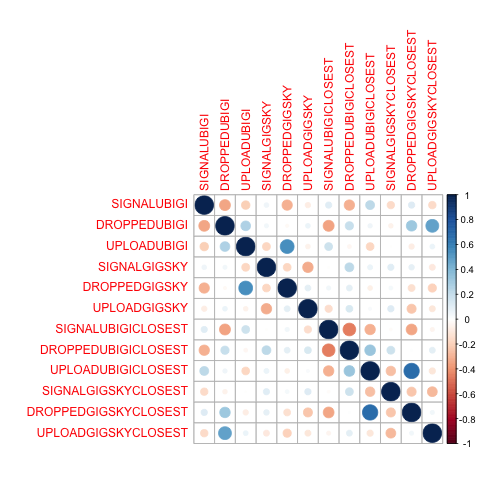}}
	\caption{Correlations between metrics in current and closest location.}
\label{distancecorrelation}
\end{figure}

Figure~\ref{correlation} shows the correlations across the metrics
and latitude and longitude coordinates. To more explicitly measure
the correlations to a specific location, we also compare the
metrics for the current location with its nearest location in
Figure~\ref{distancecorrelation}.

It is interesting to note the low correlation between the same metric for one provider
and the other. For instance getting a good signal with one provider does not mean
you will get a good signal with the other. This again strengthens the hypothesis
that selecting a different provider in a different location can lead to
improvements. Furthermore, there are significant correlations between metrics
in nearby locations, and thus there is an opportunity to predict metrics 
from observed metrics in the vicinity.

% !TEX root = aim.tex

\section{The case for a spectrum market}
\label{sec:market}

\subsection{Benefits of a spectrum market}
\label{sec:market_benefits}
Experimental observations for the simple scenario described in Sec.~\ref{sec:motivation} show (a) that a strong signal from one provider may very well coexist with a weak signal from the other provider, and vice versa; and (b) the QoE metrics for a given provider are correlated across nearby locations.  We conclude from (a) that there is potential value in a user being able to switch from one provider to another, but we also note that the inherent randomness in the wireless channel and environment means that the strength of the signal received from a selected provider may vary abruptly from instant to instant.  Given the time and delay in switching from one provider to another, it follows that a user agent should make the decision to switch based on an aggregate measure (such as the mean) QoE on the two provider networks, rather than try to keep up with instantaneous changes on the two networks.  Moreover, (b) suggests that such aggregate measures of QoE ought to be predictable for a selected provider network, based on past observations.  Together, (a) and (b) therefore imply that a user agent can improve the user's QoE by judiciously switching between providers.  We propose to support and enable such switching between providers through an \emph{online market} for short-term leases of bandwidth to user agents from multiple network operators.  The details of the implementation of such a spectrum market are described in Sec.~\ref{sec:implementation}.  In the present section, we will define an idealized and abstract version of such a spectrum market.  Doing so will allow us to formulate mathematically the provider selection problem.

\subsection{Abstract definition of a spectrum market}
\label{sec:market_defn}
Our abstract definition of a spectrum market comprises the following:
\begin{itemize}
\item A set of $k$ providers, each of whom advertises a certain block of bandwidth to lease for an advertised duration at an advertised price, all advertised information being available online to all user agents;
\item A set of user agents, each of whom may lease an advertised offering from a provider for a specified time duration at a specified price;
\item A mechanism to process, authenticate, and transfer the payments from user agents to providers for leased bandwidth.
\end{itemize}
For our purposes at the present time, it is sufficient to restrict our attention to the set of providers, and consider the actions of a single user agent trying to select a provider at time $t$ for the time duration beginning immediately and continuing until time $t+1$. Note that we are leaving the definitions of these time instants intentionally vague at this time for maximum generalizability later.

\section{The provider selection problem}
\label{sec:provider_selection}

\subsection{Terminology}
Let us restate the scenario described in Sec.~\ref{sec:market_defn} with new terminology: an \emph{agent} (on the user's device) interacts with the \emph{environment} (the rest of the devices and all providers, across all bands available on the local spectrum market) via an \emph{action} that takes the form of a purchase (strictly speaking, a lease) of bandwidth on a particular band from a particular provider for a specific duration of time.  

\subsubsection{Price setters and price takers}
We assume that the only information that the agent can access regarding the environment is the advertised ``sales catalog'' of bandwidth offerings from the various providers, together with pricing information (which we will collectively call the environmental \emph{situation} in what follows).  This pricing information may be the actual prices per unit of bandwidth for each bandwidth offering as \emph{set} by the appropriate provider. In this case, the agents are price takers. Note, however, that this does not mean that the price for a given bandwidth offering by a given provider is fixed and unchanging. On the contrary, the provider may implement congestion-based pricing (based on a proprietary algorithm that it does not disclose to other providers or to any user agent) that raises the price per unit of bandwidth if the number of users it is serving is high. In this case, the user agents indirectly influence the new prices through the congestion-pricing scheme used by the providers, but we still consider the agents to be price takers because they have no direct influence on the price.

However, the ``pricing'' information in the above environmental situation could also take the form of rules for a set of bandwidth auctions.  This set of bandwidth auctions may be designed in many ways, ranging from one bandwidth auction per provider, with the commodity/commodities in a given auction being the bandwidth offering(s) from that provider, to a single auction for all bandwidth offerings from all providers.  In the former auction arrangement, each provider only auctions off its spectrum amongst those agents who select this provider.  In the latter auction arrangement, agents bid against each other for the joint bundle of a provider and its bandwidth offering.  In either auction setup, the agents are price setters, because they directly influence the prices for bandwidth.

\subsubsection{States and contexts}
The environment may respond to the action of a given agent by changing its \emph{state}, which may in turn change the environmental situation that is the only information about the changed environment that the user agent has access to.  In this case, we call the environmental situation the \emph{state signal} and view it as a representation of the state of the environment, where we restrict the use of the term \emph{state} to properties or attributes of the environment that \emph{can be affected by the actions of an agent}.  On the other hand, if the information about the environment that is conveyed by the situation accessed by the agent is never modified by any agent action, then this situation is termed a \emph{context} instead.  

\subsection{The Reward hypothesis}
\label{sec:reward_principle}
Regardless of whether the environment changes state in response to an agent action, the environment always computes a scalar \emph{reward} (the quality of service, or QoS, for the transmissions made or received by the agent over the bandwidth purchased by its action during the time period specified in its action).  We model the goal of the user agent as that of selecting actions at each time step so as to maximize the \emph{cumulative} (i.e., summed over multiple time steps) expected reward in the long run.  This is called the \emph{Reward hypothesis}~\cite[Sec.~3.2]{sutton2018} in the literature.  Since the cumulative discounted reward, being an aggregate QoS metric, is also a measure of the value or \emph{utility} that the user gets for the actions of its agent, the \emph{Reward hypothesis} also corresponds to the principle of \emph{Expected Utility maximization}~\cite{vonneumann1944}.\footnote{Although humans do not strictly follow this principle~\cite{kahneman1979}, the user agent is not human and may be assumed to follow it.}  The distinction between \emph{state} and \emph{context} is important when it comes to the kinds of algorithms that the agent may use to determine its actions to optimize its cumulative reward, as we shall see below. For brevity, the mathematical formulation of the provider selection problem that we shall give below is for states, as it is general enough to account for context as well.

\subsection{MDP for a single user agent}
At each time step $t$, the agent takes an action $A_t$ drawn from a probability distribution (called a \emph{policy}) that depends only on the present state $S_t$ and not on the past history of states and/or actions taken to arrive at the present state.  Moreover, given $S_t=s$ and action $A_t=a$, the environment changes its state to $S_{t+1}$ and computes a reward $R_{t+1}$ according to a joint conditional probability distribution on $(S_{t+1}, R_{t+1})$ which depends only on $(s,a)$ and not the history of states and/or actions prior to time $t$.  The above defines a Markov Decision Process (MDP). Note that the action may be a selection of a particular bandwidth offering from a particular provider, and may also include a bid price for that bandwidth offering if the latter is auctioned instead of sold for a fixed price.\footnote{For reasons of allocation speed and simplicity of analysis, we assume that bandwidth auctions have only a single round of bidding, as opposed to multiple rounds.}  For the situations that we shall study in the following sections, the states will typically be either apps or pairs of apps and prices, for which it is reasonable to expect a lack of dependence of the above probability distributions on states and actions other than the most recent ones. 

\subsection{Provider selection as a Bandit problem}
\label{sec:bandit}
The provider selection problem for a single user agent is an RL problem (as per the Reward hypothesis above) wherein the user agent, without knowing the joint conditional probability distribution for the above MDP, has to find a policy that maximizes the expected value of the cumulative discounted reward 
\[
	\mathbb{E}\left[\sum_{u=0}^\infty \gamma^u R_{t+1+u} \,|\, S_1 = s_1\right]
\] 
starting from some state $s_1$ at time $t=1$, where $0 < \gamma < 1$ and $\gamma$ is a discount factor to ensure that the infinite-horizon cumulative discounted reward is finite if the rewards are bounded.  For certain forms of the conditional probability distribution on $(S_{t+1}, R_{t+1})$ given $(S_t,A_t)$, some of which we will discuss below, this general RL problem turns out to be one of a class of well-studied problems called \emph{Bandit} problems.\footnote{The original Bandit problem~\cite{robbins1952} was for a single unchanging context, not a state, but subsequent generalizations to account for state have reduced the distinction between Bandit and RL problems, especially when the underlying dynamics of the system are described by an MDP.} When we only have context (which does not change in response to an agent's actions), the algorithms to determine such optimal actions are the topic of \emph{Associative Search}, also called \emph{Contextual Bandits}~\cite[Chap.~18]{lattimore2020}.

\subsection{A learning approach}
Recall that the agent has access to just the information in the state or context regarding the environment, and it receives a reward for each action it takes.  The Reward hypothesis suggests that the way for the agent to maximize its cumulative reward is to \emph{learn} from its interactions with the environment the actions to take when in a given state or context in order to maximize not necessarily the immediate reward but the long-term cumulative reward.  Note, in particular, the flexibility and robustness of such a learning approach compared to a rule-based approach with a predefined rule or set of rules, which will always take the same action when faced with the same situation, and may not know how to react in response to a situation that is not covered by the rule.

In the next section, we present a brief review of approaches to solve the multi-armed bandit problem.

\section{Review of Multi-armed bandit problems}
\label{sec:mab_review}

\subsection{Contextual $k$-armed bandit}
\label{sec:c2ab}
We simplify the provider selection problem description by stipulating that as soon as the human user launches an app (for brevity, making a call will also be considered ``launching an app'' in what follows), the user agent instantaneously selects one of, say, $k$ SIMs or eSIMs, and enables the selected SIM/eSIM if it was not enabled already\footnote{Actually, the experimentally observed time between enabling a SIM/eSIM and being able to use it can be a few seconds (see Sec.~\ref{sec:esim}), but we will ignore this in our modeling.}.

The only \emph{action} that the user agent takes is to select the SIM/eSIM (i.e., provider) to enable next.  The \emph{context} (not changeable by the agent's action) is the app that was launched by the human user.  The \emph{reward} that the agent receives for its action is the QoE corresponding to the context (i.e., app launched on the selected provider).  Note that owing to randomness on the channel between the device and the base stations serving it, the reward is a random variable.  

In short, the agent is faced repeatedly with a choice of $k$ different actions (each corresponding to the selection of a different network provider) at each time step.  Each time step corresponds to a specific context (the app launched by the human user).  Note that the time steps do not need to be uniformly spaced.  Following the action, the agent receives a reward drawn from a probability distribution that depends on the context\footnote{A reward definition will be introduced in Sec.~\ref{sec:reward2}.}.  If we view the choice of the $k$ different actions as that of ``pulling one of $k$ levers,'' then the above is precisely the description of a \emph{contextual $k$-armed bandit}. A survey of contextual multi-armed bandits may be found in~\cite{zhou2016survey}.

The easiest way to approach a contextual $k$-armed bandit with, say, $n$ values of the context (each corresponding to a different app launched by the user) is to simply apply a (non-contextual) $k$-armed bandit separately to each of the $n$ contexts.  In other words, we ignore all relationships between the different contexts and completely decouple the bandit problems between the different contexts.  For the present scenario, this is equivalent to saying that we find, separately, the action selection rule to maximize the cumulative reward over all time steps when each particular app was launched.  Thus, in the following analysis, we will discuss only the action selection rule for the (non-contextual) $k$-armed bandit problem corresponding to a specific app.

We start with some notation.  Fix a specific app $s \in \{1,\dots,n\}$, and assume this context is unchanged in what follows. Let the time steps be numbered $1,2,\dots$.  Let the action taken by the agent at time $t$ (i.e., the label of the selected provider) in context $s$ be denoted $A_t(s) \in \{1,\dots,k\}$.  Then the simplest action selection rule is~\cite[Sec.~2.2]{sutton2018}:
\begin{equation}
	A_t(s) = \arg\max_a Q_t(s, a),
	\label{eq:avs}
\end{equation}
where for each $a \in \{1,\dots,k\}$ and $t=1,2,\dots$, the estimated action value function $Q_t(s, a)$ is defined by the arithmetic average of the rewards for action $a \in \{1,\dots,k\}$ upto and including time $t-1$:
\begin{equation}
	Q_t(s, a) = \begin{cases}
	\frac{\sum_{i=1}^{t-1} R_i 1_{\{a\}}(A_i(s))}{N_{t-1}(s, a)}, & N_{t-1}(s, a) > 0, \\
	0, & N_{t-1}(s, a) = 0,
	\end{cases}
	\label{eq:Qta}
\end{equation}
where for any $t=1,2,\dots$, $R_t = R^{(A_t(s))}(s)$ is the reward for taking action $A_t(s)$ at time step $t$ in context $s$, 
\begin{equation}
	N_{t-1}(s, a) = \begin{cases}
	0, & t = 1, \\
	\sum_{i=1}^{t-1} 1_{\{a\}}(A_i(s)), & t = 2, 3, \dots,
	\end{cases}
	\label{eq:Nta}
\end{equation}
is the number of times action $a$ was selected by the agent upto and including time $t-1$ when the context is $s$, and for any set $\mathcal{S}$, the indicator function $1_{\mathcal{S}}(\cdot)$ is defined by
\[
	1_{\mathcal{S}}(x) = \begin{cases}
		1, & \text{if } x \in \mathcal{S}, \\
		0, & \text{otherwise}.
	\end{cases}
\]
If the maximizing argument in~\eqref{eq:avs} is not unique then $A_t(s)$ is chosen from amongst the maximizing arguments at random.  In fact, to encourage exploration versus mere exploitation, we may select the action according to~\eqref{eq:avs} (breaking ties randomly as described) with probability $1 - \epsilon$ for some small $\epsilon$, say, while selecting a random action for $A_t(s)$ with probability $\epsilon$~\cite[Sec.~2.4]{sutton2018}.

Variations of the above selection rule may be defined, where the averaging in~\eqref{eq:Qta} is performed not over the entire history $1, 2, \dots, t-1$ but only the moving window of the last $w$ values at time steps $t-w, t-w+1, \dots, t-1$.  Alternatively,~\eqref{eq:Qta} may be replaced by exponential averaging with exponential smoothing coefficient $\alpha \in (0, 1)$ over the values $R_i$ for $i = 1,\dots,t-1$ where $1_{\{a\}}(A_i(s))=1$.

\subsection{Introducing state}
\label{sec:state}
Although it may appear that the launched app cannot be affected by the agent's action of provider selection and must therefore be part of the context rather than a state (recall that a state is one that can be changed by the agent's action), it is possible to redefine the above scenario to introduce a state into the problem formulation, as follows:  

Suppose we have $n$ apps, labeled $1, \dots, n$, and $k$ providers. Define the $k$ discrete-valued stochastic processes $\{S^{(i)}_t\}_{t=1}^\infty$, $i=1,\dots,k$, where $S^{(i)}_t \in \{1,\dots,n\}$, $i=1,\dots,k$ with the following dynamics: when the agent takes action $A_t = a \in \{1,\dots,k\}$ at time step $t$, the stochastic process $S^{(i)}_t$ does not transition at this time step for any $i \neq a$: $S^{(i)}_{t+1} = S^{(i)}_t$, whereas $S^{(a)}_t$ makes a transition to $S^{(a)}_{t+1}$ according to a Markov chain transition probability matrix $\bm{P}^{(a)} = [p^{(a)}_{s,s'}]_{1\leq s,s' \leq n}$, where for any $s = 1,\dots,n$ and $s' = 1,\dots,n$,
\[
	p^{(a)}_{s,s'} = \mathbb{P}\{S^{(a)}_{t+1} = s' \,|\, S^{(a)}_t = s\}, \quad a \in \{1,\dots,k\}.
\]

In other words, we are now modeling the next app, i.e., the ``next'' value $S^{(a)}_{t+1}$ for the selected provider $a \in \{1,\dots,k\}$, as being launched just \emph{after} the selection of the provider.  This is in contrast to the modeling in Sec.~\ref{sec:c2ab}, where we modeled the selection of the provider as occurring just after the launch of the next app.  The new formulation also has the benefit of accounting for the dynamics of app transitions (i.e., the behavior of the human user) instead of decoupling all the apps and treating them separately from one another as we did before in the contextual $k$-armed bandit problem.  

From the above description, it is clear that the vector stochastic process $\{\bm{S}_t = (S^{(1)}_t, \dots, S^{(k)}_t)\}_{t=1}^\infty$ is a \emph{state} of the environment (where the environment comprises the other users and all the providers), since it is changed by the user agent's action.  Let $R_{t+1} = R^{(a)}(s)$ be the reward associated with the action $A_t = a \in \{1,\dots,k\}$ and the launched app $S^{(a)}_{t+1} = s \in \{1,\dots,n\}$ on provider $a$.

Following the guidelines of RL, our goal is to identify a policy that at time step $t$, selects action $A_t$ to maximize the expected value of the discounted cumulative reward $G_t = \sum_{u=0}^\infty \gamma^u R_{t+1+u}$, where $\gamma \in (0, 1)$ is a discount factor.  

This new formulation of the provider selection problem does not have consistent nomenclature: it is most often simply called the $k$-armed bandit problem (omitting ``contextual'')~\cite{duff1995}.  It is also sometimes called the Bayesian bandit problem~\cite[Chap.~35]{lattimore2020} and sometimes (somewhat misleadingly) even called the nonstationary Bandit problem~\cite[Sec.~31.4]{lattimore2020}.  At the same time, the presence of states makes the problem amenable to attack by RL methods.  We shall discuss both approaches to solving the problem below.

\subsubsection{Direct RL approach via Q-learning}
\label{sec:l1rl}
From the Bellman equations, the optimal action in state $\bm{S}_t$ is given by
\[
	A_t = \arg\max_a q_*(\bm{S}_t, a),
\]
where the \emph{action-value function} (or $n^k \times k$ table) $q_*(\bm{s}, a)$ is defined as~\cite[eqn.~(3.20)]{sutton2018}
\begin{align}
	&q_*(\bm{s}, a) = \notag \\ 
	&\mathbb{E}\left[R^{(a)}(S^{(a)}_{t+1}) + \gamma \max_{a'} q_*(\bm{S}_{t+1}, a') \bigm | \bm{S}_t = \bm{s}, A_t = a\right],
	\label{eq:l1avf} \\
	&\bm{s} \in \{1,\dots, n\}^k, \quad a \in \{1,\dots,k\}. \notag
\end{align}

We could perform the same kind of sampling and averaging for each state-action pair $\bm{S}_t, a)$ as in Sec.~\ref{sec:c2ab} for each context-action pair, except that the averaging would be over the discounted cumulative rewards $G_t$ rather than the raw rewards $R_t$.  This is an example of a Monte Carlo method~\cite[Sec.~5.2]{sutton2018}.  We choose not to employ RL Monte Carlo methods (in contrast to contextual bandit Monte Carlo methods, which we do employ), because the former require averaging over distinct \emph{episodes}, where the sequence of states ends with a so-called \emph{terminal state} representing an exit or end to a session.  Such episodes are not clearly defined for our use case.

Instead, we use the Q-learning method~\cite[Sec.~6.5]{sutton2018} to estimate $q_*(\bm{s}, a)$ by iteratively updating at each time step as follows:
\begin{align}
	&q_*(\bm{s}, a) \leftarrow \notag \\
	&q_*(\bm{s}, a) + \alpha [R_{t+1} + \gamma \max_{a'} q_*(\bm{s}', a') - q_*(\bm{s}, a)],	
	\label{eq:L1QL}
\end{align}
where $\bm{s}$ is the present state, $\bm{s}'$ and $R_{t+1}$ are respectively the next state caused by, and reward associated with, the agent action $a$ in the state $\bm{s}$, and $\alpha \in (0, 1)$ is an exponential smoothing coefficient.

Note that by the definition of the state vector, the transition from state $\bm{s}$ to $\bm{s}'$ under the action $a$ only changes the $a$th entry of $\bm{s}'$ relative to $\bm{s}$.  Let $s$ be the $a$th entry of $\bm{s}$. Then the updates in~\eqref{eq:L1QL} may be seen as applying to the $n \times k$ table $\tilde{q}_*(s, a)$ instead of to $q_*(\bm{s}, a)$, where $\tilde{q}_*(s, a)$ is just the function $q_*(\bm{s}, a)$ with all entries of $\bm{s}$ fixed except for the $a$th entry.  Thus we need maintain and update only an $n \times k$ table instead of an $n^k \times k$ table.

\subsubsection{Exact and approximate (RL) approaches to the $k$-armed bandit problem}
\label{sec:duff}
It is a remarkable result~\cite[Thm.~35.9]{lattimore2020} that if the transition probability matrices $\bm{P}^{(a)}$ are known, and the expected reward $\mathbb{E}[R_{t+1} | S_{t+1}, S_t, A_t]$ is a deterministic function of $S_{t+1}$, then $\mathbb{E}[\sum_{u=0}^\infty \gamma^u R_{t+1+u}]$ can be maximized by a deterministic policy of the form $A_t(\bm{S}_t) = \arg\max_{1 \leq a \leq k} g_a(S_t)$, where $g_a(\cdot)$ is called the Gittins index and can be computed from the known transition probability matrix $\bm{P}^{(a)}$ by an iterative algorithm like the Varaiya-Walrand-Byukkoc algorithm~\cite[Sec.~35.5]{lattimore2020}.  

Thus the optimal policy for the provider selection problem is available if we can compute the Gittins indices.  However, this would require knowledge of the app transition probability matrices, which is unavailable to the agent.

Duff~\cite{duff1995} proposed instead to apply Q-learning to approximate the calculation of the Gittins indices by updating not one but two new action-value functions (each an $n \times k \times n$ table) at each training step, where one of these action-value functions is the Gittins index to be approximated, and the action at that training step is from the softmax distribution over this action-value function.  Note that we maintain and update $2k\,n^2$ values over time, so the numerical complexity is higher than when we directly apply Q-learning (maintaining and updating only $k\,n$ values over time) to maximize the discounted cumulative reward as in Sec.~\ref{sec:l1rl}. Therefore we will not use this approach in the present paper.  However, we mention it here for future reference, as it can be applied to the provider selection problem in the general form of the spectrum market, as discussed in Sec.~\ref{sec:dualspeedrestless}.

\subsection{Quality of Experience (QoE)}
\label{sec:QoE}
So far we have not given an expression or definition of a particular reward function for the two modeling scenarios in Sec.~\ref{sec:c2ab} and Sec.~\ref{sec:l1rl}. In either scenario, the action $A_t$ of selecting a provider at time step $t$, immediately preceded or followed by launching an app $s \in \{1,\dots,n\}$, associates the reward $R_{t+1}$ with this action.  Before we can define the form of the reward $R_{t+1}$, we need to define the \emph{Quality of Experience} (QoE) which is an important component of the reward.

We denote by $\mathtt{QoE}^{(a)}_{t+1}(s)$ the QoE to the user on the selected provider $a$ over the session that begins when the app $s$ is launched and ends when the app is closed or another app is launched, i.e., at time $t+2$.  For example, $\mathtt{QoE}^{(a)}_{t+1}(s)$ could be the throughput if $s$ is a certain app, and could be latency if $s$ is a different app. Note that the QoE is a random variable. 

In order to be able to compare and accumulate rewards corresponding to different apps being launched, we assume that for each launched app $s$, $\mathtt{QoE}^{(a)}_{t+1}(s) \in \{1, 2, \dots, 10\}$ is the \emph{decile rank} of the observed QoE relative to its own (i.e., same app, and same provider) probability distribution.  The probability distribution is unknown, but is estimated from usage history.

An example of a reward function and its dependence on the QoE is given in Sec.~\ref{sec:reward2}.

\section{Market with fixed prices and equal allocations}
\label{sec:level2}
In this section, we formulate the provider selection problem for the scenario(s) that we shall then evaluate via simulation and experiment.  At any time step, we will allow the providers to change the prices they charge for bandwidth, provided that these prices then stay fixed until the next time step.  We also assume that the actions of the agents on the user devices in the network do not directly determine the prices charged by the providers, although they may indirectly determine these prices, as discussed in Sec.~\ref{sec:reward_principle}.  

We will restrict ourselves to the scenario where at each time step $t$ a user agent's action $A_t$ is merely that of selecting a provider $a \in \{1,2,\dots,k\}$ to purchase bandwidth from at the present time step $t$, with this bandwidth to be used for transmissions over the next \emph{epoch}, which is defined as the time interval starting just after time step $t$ and ending at time step $t+1$. This represents the implementation of the spectrum market in Sec.~\ref{sec:implementation}, where the Blockchain used by the provider is set up to certify a certain number of user purchases for a block of bandwidth during a specified interval.  In other words, the provider's offerings on the spectrum market are advertised in the following way, say: ``\$5 for 20~MHz of bandwidth, will accept up to 3 contracts for the hour starting 10am.'' For this example, the provider stands to collect \$5 each from 1, 2, or 3 user agents, who may select this provider at any time between 10am and 11am.  Moreover, the provider divides the available bandwidth evenly between all users who select this provider during this time period. If only one user agent selects this provider, it pays \$5 to the provider and receives the entire 20~MHz; if, however, a few minutes later, a second user agent also selects this provider, it pays \$5 to the provider, and now both user agents receive only 10~MHz each for the rest of the hour, and so on.  In short, since no user agent knows how many other user agents have selected a given provider, the bandwidth that an agent will receive on selecting a given provider is unknown to the agent.  

\subsection{Contextual multi-armed bandit}
\label{sec:cmab}
Recall that the prices charged by the providers to the agents for selecting these providers, and these prices cannot be changed by the agents' actions, although they may be changed by the providers from one time step to the next.  Clearly, this allows us to consider the set of prices charged by all providers as part of the context.  As before, the app launched at any time step is also part of the context.  From the perspective of a given user agent, the context at time $t$ may therefore be defined as the pair $(s_t, \bm{p}_t)$, where $s_t$ is the app launched at time step $t$ and $\bm{p}_t = (p^{(1)}_t,\dots,p^{(k)}_t)$ is the set of prices charged by the providers for connecting to their networks.

The above discussion makes it clear that we can apply the contextual multi-armed bandit algorithm of Sec.~\ref{sec:c2ab}.  We will adopt the simple approach of decoupling the contexts as before and considering them as separate non-contextual multi-armed bandit problems.  Thus the action selection rule is that of~\eqref{eq:avs}, where the estimated action value function is given by~\eqref{eq:Qta}.  We call this action selection rule \emph{ExpectedUtility}, noting that the expectation operation is not necessarily just the arithmetic mean as defined in~\eqref{eq:Qta}, but may also be a moving average over a finite window, or exponential smoothing, as discussed in Sec.~\ref{sec:c2ab}.  Recall that the context space is now no longer just the launched app but the pair $(s_t, \bm{p}_t)$ defined above.  Thus, if there are $k=2$ providers, each randomly choosing one of $m=2$ prices to set at a given time step, then $\bm{p}_t = (p^{(1)}_t,\dots,p^{(k)}_t)$ takes one of $k^m=4$ values.  If the user launches one of $n=2$ apps, then $s_t$ takes one of $n=2$ values, and then the context $(s_t, \bm{p}_t)$ can take one of $n\,k^m = 8$ values.  The simple decoupling approach would indicate that for each value of the context, there needs to be a separate estimated action value function~\eqref{eq:Qta}.  However, we note that for a given agent action (provider selection) $a$ at time step $t$, the only entry of $\bm{p}_t$ that matters is $p^{(a)}_t$.  It follows that we only need separate action value function estimates~\eqref{eq:Qta} for each of the $n\,m$ values of the reduced context $(s_t, p^{(a)}_t)$.  

\subsection{Q-learning solution}
\label{sec:l2rl}
Just as we did earlier in Sec.~\ref{sec:l1rl}, it is possible to convert the context $(s_t, \bm{p}_t)$ into a state $(\bm{s}_t, \bm{p}_t)$ that can be changed by the agents' actions just by modeling both the app-transitions and the pricing-transitions as transitions to different stochastic processes that occur immediately after the agent takes an action at a given time step.  This also has the virtue of accounting for the dynamics of app transitions (i.e., human user behavior) as well as pricing transitions (i.e., provider behavior), instead of decoupling all values of $(\bm{s}_t, \bm{p}_t)$ from one another and treating them separately.  

Moreover, if the agent takes the action $A_t=a$ of selecting provider $a \in \{1,2,\dots,k\}$, as a result of which the state changes from $(\bm{s}_t, \bm{p}_t)$ to $(\bm{s}_{t+1}, \bm{p}_{t+1})$, the only components of the state that have actually changed are the $a$th entries of $\bm{s}_t$ and $\bm{p}_t$ respectively.  In other words, we can work with the reduced state $(s^{(a)}_t, p^{(a)}_t)$, which has only $n\,m$ values.  Thus, in the same way as described in Sec.~\ref{sec:l1rl}, we can apply Q-learning by updating a reduced $n\,m \times k$ table $\tilde{q}_*((s, p), a)$ as per~\eqref{eq:L1QL} at each time step.  Also, this table can be augmented with probabilities $\pi_t((s, p), a)$ calculated from~\eqref{eq:pi1} in order to perform action selection according to~\eqref{eq:L1QLc} as discussed in Sec.~\ref{sec:l1rl}.  

Finally, we note that although the Q-learning approximation to the Gittins Indices proposed in~\cite{duff1995} and discussed in Sec.~\ref{sec:duff} is also applicable here, we do not pursue it further for the same reason as in Sec.~\ref{sec:duff}, namely the higher numerical complexity compared to direct Q-learning applied to maximize the cumulative discounted reward.

\subsection{Reward function}
\label{sec:reward2}
We define the reward as the ``value for money'' where the ``value'' is the QoE:
\begin{equation}
	R_{t+1} = 
	\frac{\mathtt{QoE}^{(a)}_{t+1}(s)}{\mathtt{PlanPrice}^{(a)}_t}, 
	\label{eq:l2r1}
\end{equation}
and $\mathtt{PlanPrice}^{(a)}_t$ is the price charged by the selected provider $a$ to the user agent for joining this provider's network.
% !TEX root = aim.tex

\section{Simulation}\label{sec:simulation}
The purpose of the simulations is to quantify the
benefits of learning algorithms in a mobile
network with dynamically-priced bandwidth
contracts.

More specifically, we study the effectiveness
of mobile network provider selection based on
experience quality of service, called utility here.
When a selection is made historical experiences
as well as current costs are known. We also
assume that each UE knows its own demand but
not the demand of other UEs. Demand of other
UEs need to be learned or inferred and cannot be 
directly obtained or queried. 

One reason for this is that the network providers
may not be willing to share this information,
another is that it depends very much on the
demand and preferences of a UE how it impacts other UEs.

The simulator computes the effective bandwidth and
QoE for the UE using basic utility functions and
SINR estimates based on positions of UEs, Base
Stations and resource contention when multiple UEs
connected to the same Base Station contend
for spectrum resources.

Since the delivered performance is computed,
it is more deterministic than with a real-world
scenario where measurements may have non-negligible
variance, so we also complement the simulations with
experiment with real phones and networks
in Sec.~\ref{sec:experiment}.

There are two types of UEs in the experiment:
Background UEs that are simply placed in the
simulation grid to inject contention, also
referred to simply as UEs, and UEs that
are allowed to make network selection
decisions dynamically, and that we capture
performance from, referred to as Device under
Test (DUT).

We also simulate the complex problem of competing
DUTS, or competing agents. It is easy to imagine
algorithms where each agent makes the same decision
and thus causes oscillation or lockstep behavior and load
imbalance. We also hypothesize
that varying demand across agents can be exploited
to put them on different optimal providers and thereby
yield a higher aggregate utility or QoE. This
scenario also makes it clear that not only
does a DUT not know the future demand of other DUTs, but it does not
even know which provider the other DUTs will pick,
and thereby potentially causing contention impacting the DUT QoE. 

\subsection{Setup}
We implemented a discrete even simulator based on
the PyLTEs framework~\cite{slabicki2014}. PyLTEs was extended to
add dynamic bandwidth pricing, UE
resource contention, DUT network selection and competition,
stochastic app demand, Base station position offsets
across networks, a straight-path mobility model,
and utility evaluation and recording.

Two competing networks are configured where one
has a higher number of background UEs causing the
maximum throughput delivered to be lower.
The maximum throughput depends on the distance
between the base station and the UE as well
as the number of other UEs using the same base station.

The network base stations are positioned with an offset
and the DUTs start in the center and then
move in a straight path away from the center at
a random angle.

Recall the discussion in Sec.~\ref{sec:reward_principle} that the expected cumulative reward may be seen as a kind of utility for a given user agent.  Our evaluation metric is aggregate utility, a.k.a.
social welfare across all DUTs. Utility, being the expected cumulative reward, is computed
based on the maximum throughput delivered and the
demand of the currently running app on the DUT.
Apps run with a transition probability to use the
same app in the next step or switch to a new app.
Each app has an associated utility function to
compute the reward. A batch app simply computes the
reward as the maximum throughput over price, and an
interactive app sets a minimum throughput threshold
that needs to be delivered to receive the full reward
and caps the reward at that level. If that throughput
is not met a lower bar reward slightly above 0 is
delivered. For simplicity we specify that
both app types are equally likely at any given time.  Since both rewards are based on throughputs, we will for simplicity work with the throughput-to-price ratios themselves instead of converting them to their decile ranks as defined in Sec.~\ref{sec:QoE}.

The intuition is that a network selector that wants to
optimize utility (i.e., expected cumulative reward) could select a lower cost, lower
throughput network if the price is right.

Below we run simulations to investigate: optimal
history length to estimate app utility, various
combinations of fixed price and location,
and competing DUTs.

The fixed location configuration ensures that
the two networks always deliver a fixed max
throughput throughout the simulation. The fixed
price configuration similarly ensures that the networks
do not change their prices within a run.

In the competing agent setup, multiple DUTs
get to pick their preferred network in each
step without knowing the decisions of other
DUTs. DUTs train or calibrate at different
times in this case to avoid lockstep behavior.

Each simulation is run in 200 steps (using
the straight-walk mobility model described above).
For each run or iteration we compute the social welfare for
each benchmark. The costs and apps and positions
are replayed for all benchmarks. We then iterate
over the same procedure generating new app, cost
and position traces 100 times and compute statistics
with each iteration as an independent sample.

We set up 36 base stations per network in a grid with radius
1440 meters in a hexagonal layout with one network
offset in both x and y coordinates from the other.
In the fixed location case we set a step walk length
to 1 meter ensuring that the signal from the networks
does not change significantly to change the max throughput
delivered from each network during the run. For the variable
location setting the walk size is set to 20 meters
for each step, which results in the networks delivering 
different throughput over time.

A summary of the general configuration parameters used
across all simulations are shown in 
Table~\ref{T:simulation_config}.
Fixed location is achieved by setting {\it walk length} to 1,
fixed pricing is enforced by setting {\it max cost} 
equal to {\it min cost}, and finally competing agents
are achieved by setting {\it DUTs} to 3.

\begin{table*}[htbp]
        \caption{Default Simulation Configuration.}
\begin{center}
\begin{tabular}{|l|l|}
\hline
\textbf{\it General} &   \\
\textbf{Steps} & 200  \\
\textbf{Training Steps} & 30 \\
\textbf{Iterations} & 100 \\
\textbf{Networks} & 2 \\
\textbf{Base Station Cell Layout} & hexagonal \\
\textbf{Cell Radius (m)} & 1666 \\
\textbf{Walk Length (per step, m)} & 20 \\
\textbf{DUTs} & 1 \\
\textbf{Mobilty Model} & Straight path from center at random angle \\
\hline
\textbf{\it Network 1} &  \\
\textbf{Base Station Power (dBm)} & 30  \\
\textbf{UEs} & 72  \\
\textbf{Base Stations} & 36 \\
\textbf{X,Y BS Grid Offset (units of radius) } & $0.6,0.4$ \\
\textbf{Min Cost (\$)} & 1 \\
\textbf{Max Cost (\$)} & 2 \\
\hline
\textbf{\it Network 2} &  \\
\textbf{Base Station Power (dBm)} & 100  \\
\textbf{UEs} & 0  \\
\textbf{Base Stations} & 36 \\
\textbf{X,Y BS Offset} & $0,0$ \\
\textbf{Min Cost (\$)} & 9 \\
\textbf{Max Cost (\$)} & 10 \\
\hline
\textbf{\it App Demand} &  \\
\textbf{App 1 Utility Function} & interactive \\
\textbf{App 1 Threshold Demand (Mbps)} & 12 \\
\textbf{App 2 Utility Function} & batch \\
\textbf{Transition Probabilities App1$\rightarrow$App1,App1$\rightarrow$App2} & $0.5,0.5$ \\
\textbf{Transition Probabilities App2$\rightarrow$App1,App2$\rightarrow$App2} & $0.5,0.5$ \\
\hline
\end{tabular}
\label{T:simulation_config}
\end{center}
\end{table*}
Note, app demand is individually and independently sampled on each DUT
(in the competing agent case).

\subsection{Provider selection policies evaluated via simulation}
We evaluated the following provider selection policies:
\begin{enumerate}
\item \emph{ExpectedUtility}: As discussed in Sec.~\ref{sec:cmab}, this is the contextual $k$-armed ($k=2$) bandit optimal policy given by~\eqref{eq:avs} with $n=2$ contexts (apps), where the function being maximized is given by~\eqref{eq:Qta}.  
\item \emph{History}: Same as \emph{ExpectedUtility}, except that we define not $n$ contexts (one per app) but only a single context.  In other words, this is the original (non-contextual) $k$-armed bandit of~\cite{robbins1952}.
\item \emph{RL}: Recall that we have $n=2$ apps and $k=2$ providers in the simulation.  Although each provider can set one of $m=2$ prices in the simulation, we simplify the state space for the Q-learning RL solution by using the smaller $n \times k$ (app, provider) action-value table of Sec.~\ref{sec:l1rl} instead of the $n\,m \times k$ ((app, price), provider) table of Sec.~\ref{sec:l2rl}. Specifically, we select the provider at time $t$ according to $A_t = \arg\max_{a \in \{1,\dots,k\}} \tilde{q}_*(S_t^{(a)}, a)$, where for any $s \in \{1, 2\}$ and $a \in \{1, 2\}$, 
\begin{align}
	&\tilde{q}_*(s, a) \leftarrow \notag \\
	&\tilde{q}_*(s, a) + \alpha [R_{t+1} + \gamma \max_{a'} \tilde{q}_*(s', a') - \tilde{q}_*(s, a)],	
	\label{eq:rlsim}
\end{align}
where $\alpha = 0.2$ and $\gamma = 0.7$.
\item \emph{LowestPrice}: This is a baseline policy for comparison purposes, where at each time step we simply select the provider charging the lower price of the two providers.
\item \emph{Random}: This is another baseline policy that is evaluated purely to serve as a comparison against \emph{History}, \emph{ExpectedUtility}, and \emph{RL}.  Here, one of the two providers is selected by tossing a fair coin at each time step, independently from one step to the next.
\end{enumerate}

\subsection{History Parameter Tuning}
We first investigate the optimal history length of
recorded throughput values for apps to best
estimate optimal utility on the two networks. Our parameter study compares the results
of using as much history as available versus only
keeping the 1-4 latest measurements. 
Pairwise t-tests were performed using 2-step rolling
averages for each iteration (50 pairs total) with
Bonferroni adjustments.
The improvement of a benchmark against a baseline is simply
computed as the difference over the baseline. 

\begin{table}[htbp]
        \caption{History Parameter Tuning p-values and average improvements in t-test against unlimited history baseline.}
\begin{center}
\begin{tabular}{|l|l|l|}
\hline
      &  \textbf{p-value} & \textbf{Improvement} \\
\hline
1-Period & $0.03$ & $0.033$ \\
2-Period & $0.008$ & $0.037$ \\
3-Period & $0.03$ & $0.033$ \\
4-Period & $0.08$ & $0.030$ \\
\hline
\end{tabular}
\label{T:history_tuning}
\end{center}
\end{table}

\begin{figure}[htbp]
        \centerline{\includegraphics[scale=0.25]{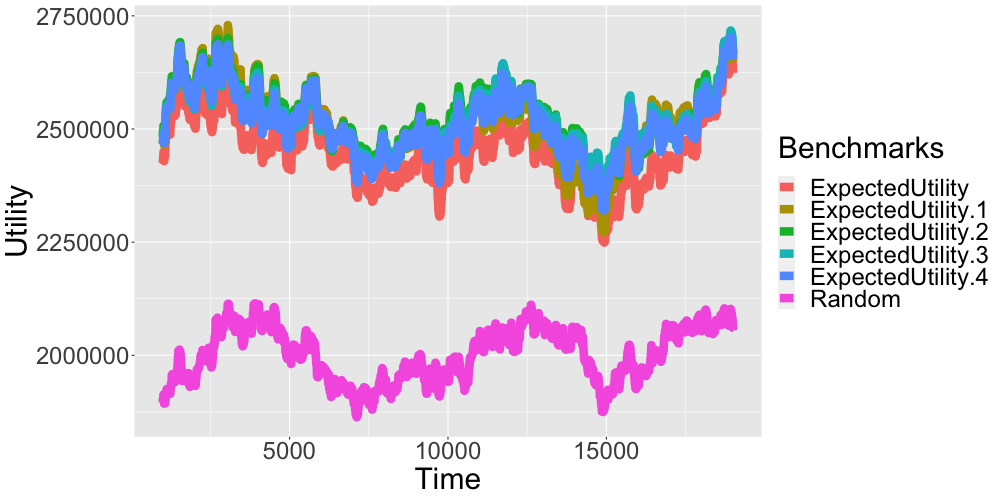}}
        \caption{History Parameter Tuning Unlimited, 1,2,3, and 4-period History.}
\label{history}
\end{figure}

Based on the results in Table~\ref{T:history_tuning} and Figure~\ref{history}, the
2-period history outperforms the other parameters slightly
and will therefore be used henceforth. As noted
above, in a real-world experiment the optimal
history lengths are likely to be longer and they also
depend on the speed of travel, as well as the
mobility model assumed. The point here
is simply that this window averaging is effective in
coping with mobility.

\subsection{Fixed Location, Fixed Price}
\begin{figure}[htbp]
        \centerline{\includegraphics[scale=0.25]{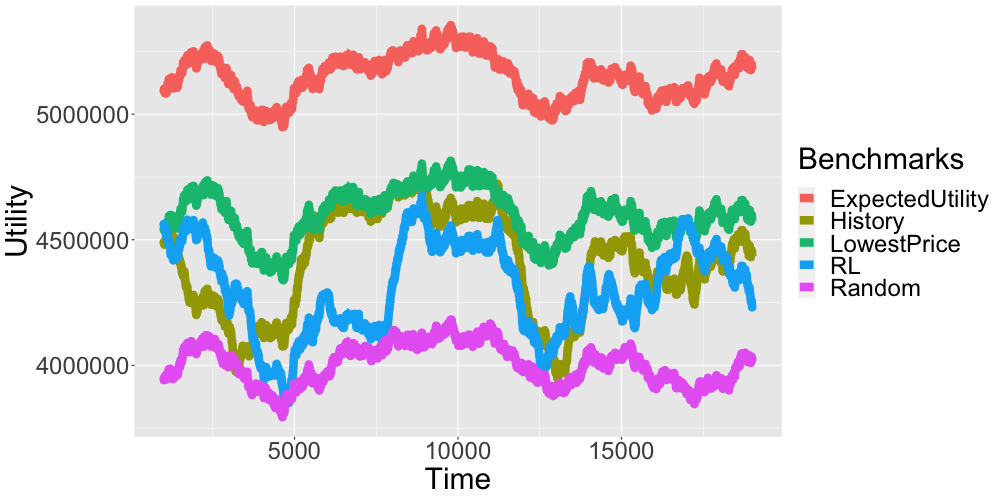}}
        \caption{Fixed Location, Fixed Price.}
\label{fixedall}
\end{figure}

The simples possible scenario is when the DUT does not move
outside of the cell and the max throughput environment does not
change (fixed location) and the prices on the two networks
do not change (fixed price). In this case the app demand
is the only variable, and the only thing learned is the
load on the networks and how well they meet the demand.
The results summarized in Table~\ref{T:fixedlocfixedprice}
and depicted in Figure~\ref{fixedall} show that the {\it ExpectedUtility}
model outperforms the other models significantly between
$14$ and $34$\% on average. Notably, the {\it History} benchmark
does poorly in comparison due to not taking the App demand into
account, i.e. not being aware of which app is causing what QoE.

\begin{table}[htbp]
        \caption{Fixed Location, Fixed Price p-values and average improvements of 2-period ExpectedUtility in t-test.}
\begin{center}
\begin{tabular}{|l|l|l|}
\hline
      &  \textbf{p-value} & \textbf{Improvement} \\
\hline
History & $2\times10^{-16}$ & $0.19$ \\
LowestPrice & $2\times10^{-16}$ & $0.14$  \\
RL & $2\times10^{-16}$ & $0.23$  \\
Random & $2\times10^{-16}$ & $0.34$  \\
\hline
\end{tabular}
\label{T:fixedlocfixedprice}
\end{center}
\end{table}

\subsection{Fixed Location, Variable Price}
Next, we study the case where the price is no longer fixed, i.e. it could change over
time even if the throughput doesn't change which provider is optimal for an app demand.
In Table~\ref{T:fixedlocvarprice} and Figure~\ref{fixedprice} we can see that {\it History} which also takes
cost into account now performs a bit better, but still worse then {\it ExpectedUtility}.
The standard {\it RL} model does worse compared to the fixed price scenario
and so does the {\it LowestPrice} benchmark that simply picks the cheapest provider. The {\it Random}
model scores slightly better in this scenario which could be interpreted as the load balancing
across the different providers is more important when the price varies too. Overall the {\it ExpectedUtility}
model performs similarly compared to the fixed price scenario and dominates all other models, showing
that price is accounted for appropriately.

\begin{table}[htbp]
        \caption{Fixed Location, Variable Price p-values and average improvements of 2-period ExpectedUtility in t-test.}
\begin{center}
\begin{tabular}{|l|l|l|}
\hline
      &  \textbf{p-value} & \textbf{Improvement} \\
\hline
History & $2\times10^{-16}$ & $0.12$ \\
LowestPrice & $2\times10^{-16}$ & $0.20$ \\
RL & $2\times10^{-16}$ & $0.30$ \\
Random & $2\times10^{-16}$ & $0.27$ \\
\hline
\end{tabular}
\label{T:fixedlocvarprice}
\end{center}
\end{table}

\begin{figure}[htbp]
        \centerline{\includegraphics[scale=0.25]{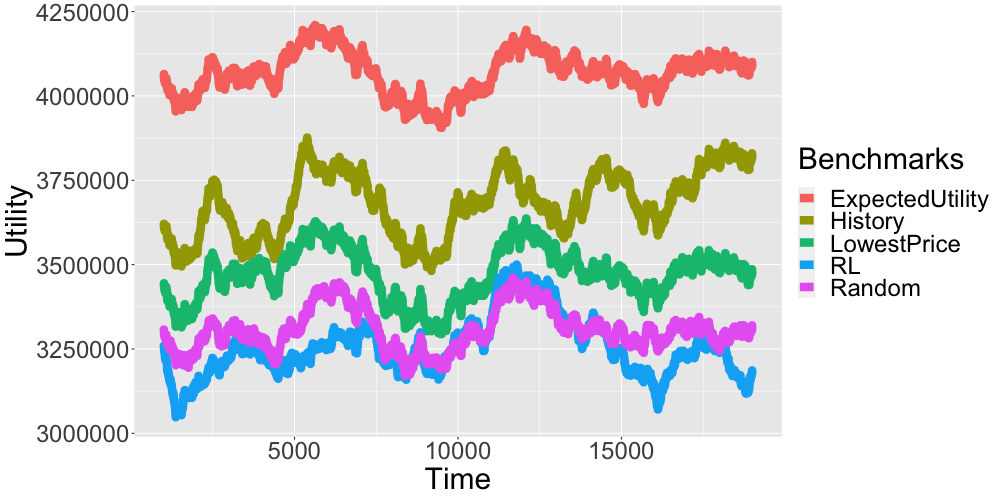}}
        \caption{Fixed Location, Variable Price.}
\label{fixedloc}
\end{figure}

\subsection{Variable Location, Fixed Price}
\begin{figure}[htbp]
        \centerline{\includegraphics[scale=0.25]{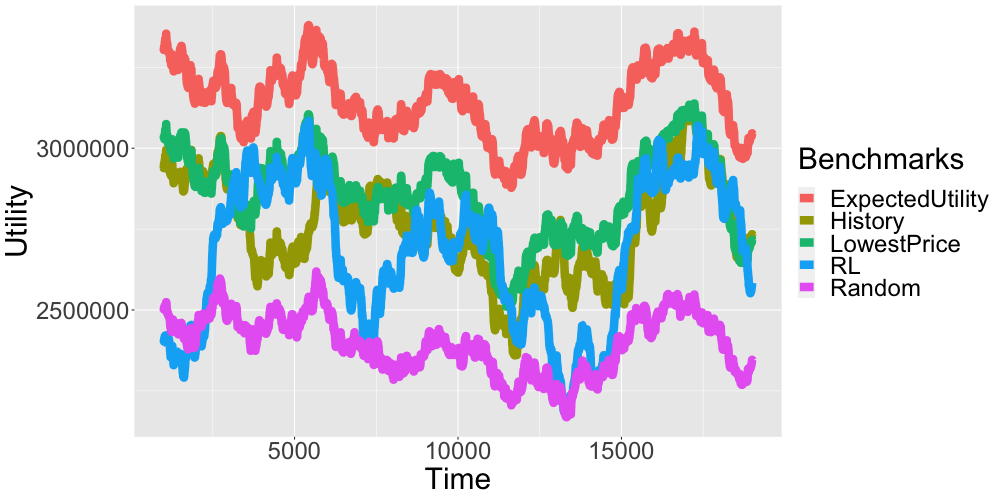}}
        \caption{Variable Location, Fixed Price.}
\label{fixedprice}
\end{figure}

Now, we vary location but fix the price over time.
Most notably from Table~\ref{T:varlocfixedprice} and Figure~\ref{fixedprice} is that
the {\it Random} allocation does very poorly with more than $40$\% worse aggregate utility
compared to {\it ExpectedUtility}. This shows that price dynamics could easily make a random allocator not taking cost into account perform
very poorly. Compared to the variable price case {\it History} also does
a bit worse, which may be explained by the fact that it doesn't take location explicitly into account in terms of a window
averaging mechanism like {\it ExpectedUtility}.

\begin{table}[htbp]
        \caption{Variable Location, Fixed Price p-values and average improvements of 2-period ExpectedUtility in t-test.}
\begin{center}
\begin{tabular}{|l|l|l|}
\hline
      &  \textbf{p-value} & \textbf{Improvement} \\
\hline
History & $1.5\times10^{-15}$ & $0.17$ \\
LowestPrice & $1.2\times10^{-8}$ & $0.12$ \\
RL & $2\times10^{-16}$ & $0.23$ \\
Random & $2\times10^{-16}$ & $0.41$ \\
\hline
\end{tabular}
\label{T:varlocfixedprice}
\end{center}
\end{table}

\subsection{Variable Location, Variable Price}
\begin{figure}[htbp]
        \centerline{\includegraphics[scale=0.25]{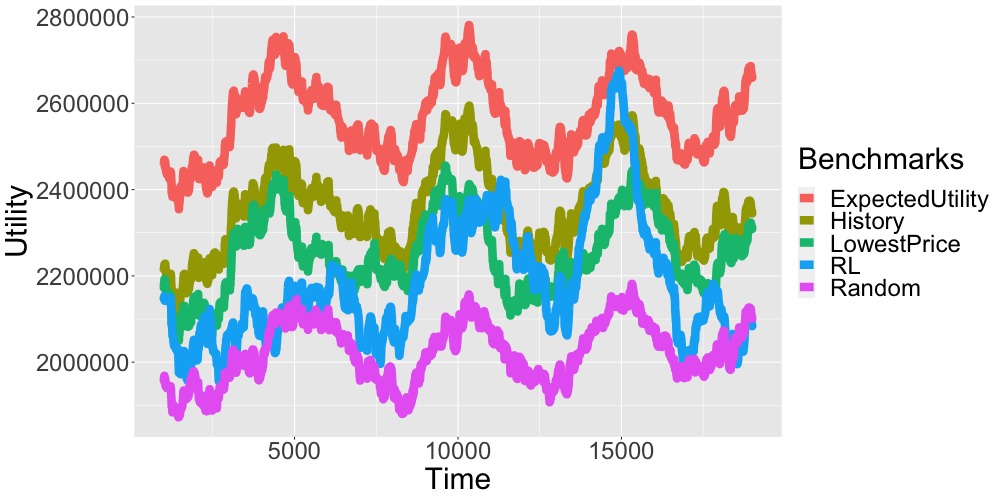}}
        \caption{Variable Location, Variable Price.}
\label{fixednone}
\end{figure}

Next, we vary both location and price, and see that {\it ExpectedUtility} maintains its
dominance over the other models in Table~\ref{T:varlocvarprice} and Figure~\ref{fixednone}.
Notably {\it LowestPrice} does poorly here, making the case that taking both price and
mobility into account is important (in addition to demand).

\begin{table}[htbp]
        \caption{Variable Location, Variable Price p-values and average improvements of 2-period ExpectedUtility in t-test.}
\begin{center}
\begin{tabular}{|l|l|l|}
\hline
      &  \textbf{p-value} & \textbf{Improvement} \\
\hline
History & $3.8\times10^{-14}$ & $0.12$ \\
LowestPrice & $2\times10^{-16}$ & $0.18$ \\
RL & $2\times10^{-16}$ & $0.23$ \\
Random & $2\times10^{-16}$ & $0.36$ \\
\hline
\end{tabular}
\label{T:varlocvarprice}
\end{center}
\end{table}

\subsection{Competing Agents}
\begin{figure}[htbp]
        \centerline{\includegraphics[scale=0.25]{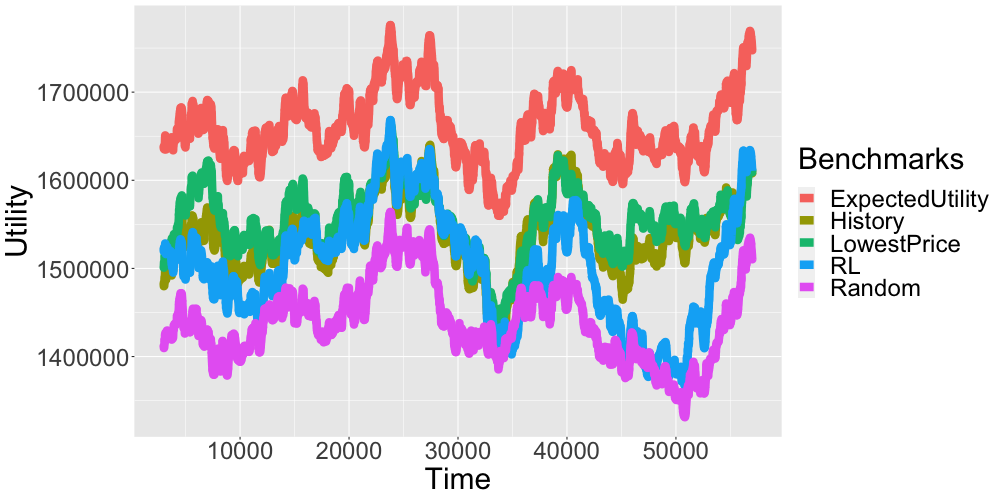}}
        \caption{3 Competing DUTs.}
\label{multi}
\end{figure}

Finally, we vary both price, location and demand and also introduce two more competing DUTs,
running the same provider selection algorithms, but with independent demands.
Surprisingly, from Table~\ref{T:multidut} and Figure~\ref{multi} the standard {\it RL} does
comparatively better although still worse than {\it ExpectedUtility}. The fact that the
{\it ExpectedUtility} improvement over {\it Random} drops from about $40$ to $20$\% could 
be explained by the fact that the decisions other DUTs are making could mislead the agent
into thinking a network is worse than it is. It is still promising that the {\it ExpectedUtility}
method does best even in this scenario.

\begin{table}[htbp]
        \caption{3 Competing Agents, p-values and average improvements of 2-period ExpectedUtility in t-test.}
\begin{center}
\begin{tabular}{|l|l|l|}
\hline
      &  \textbf{p-value} & \textbf{Improvement} \\
\hline
History & $5.9\times10^{-7}$ & $0.13$ \\
LowestPrice & $6.1\times10^{-13}$ & $0.19$  \\
RL & $0.0024$ & $0.085$ \\
Random & $1.6\times10^{-13}$ & $0.20$ \\
\hline
\end{tabular}
\label{T:multidut}
\end{center}
\end{table}

\subsection{Summary of simulation results}
We observe that in each of the simulated scenarios, the Monte Carlo algorithm for contextual multi-armed bandits, which we call \textit{ExpectedUtility}, performs better than the direct Q-learning algorithm applied to maximize the expected cumulative discounted reward, which we have called \emph{RL} above.  Given the relative lack of sophistication of the Monte Carlo algorithm~\eqref{eq:avs} compared to the RL Q-learning algorithm~\eqref{eq:rlsim}, these results may be surprising.  However, they are explainable given that the simulated scenarios are exactly contextual multi-armed bandit problems, and the Monte Carlo algorithm is, in spite of its simplicity, a state-of-the-art solution to such problems~\cite[p.~43]{sutton2018}, performing as well or better than many other algorithms including deep Q-learning (DQN)~\cite[Tab.~1]{riquelme2018}.  On the other hand, Q-learning is known to be hard to train with low training sample efficiency~\cite{tsitsiklis1996} and has been bettered in performance by other contextual multi-armed bandit methods in the literature~\cite{cannelli2020}.

% !TEX root = aim.tex

\section{Implementation Notes}\label{sec:implementation}
In this section we go into some more details on how
the proposed system has been implemented. as a prelude to
our experiments. The current
implementation relies on eSIM functionality and hence
we start off with a quick primer on eSIM. 

% !TEX root = aim.tex

\subsection{eSIM}\label{sec:esim}
Embedded Subscriber Identity Module (eSIM) or Embedded Universal
Integrated Circuit Card (eUICC) is a programmable chip embedded
in a device that allows it to connect to different
mobile networks without a physical SIM card. The eSIM specifications
are developed by the GSM Association (GSMA) and define the protocols
and components required to remotely provision a software-based
SIM card profile as part of subscribing to a mobile network service~\cite{gsma}.

An eSIM profile is typically downloaded from a Subscription Manager - Data
Preparation+ (SM-DP+) server certified by the GSMA using a QR-code containing
an activation code. The download process maps the identity of the device to 
a subscription provided by a mobile network operator.
After the profile has been downloaded it may be activated, at which point the
eSIM authenticates with and connects to a network with a matching public
land mobile network (PLMN) identifier within reach, potentially
after roaming to a supported provider. Typically only a single eSIM profile may be active at
any given time, but any number of profiles may be downloaded and be in
an inactive state on the device. After the eSIM is activated it behaves
in the exact same way as a physical SIM card, until it is deactivated by switching to
another profile or by deleting it from the device.

The time it takes to switch depends on the provider, and can be substantial if roaming
is involved. The overhead beyond the authentication process (e.g. LTE Attach) is, however, 
negligible. 

The eSIM profile contains a hash of the mobile network provider certificate 
that allows mobile apps developed by that same provider to manage the eSIM profile
in what is known as a carrier app. Providers have no access to profiles they
did not provision (i.e. they cannot download or switch to profiles they do not own).

Hence, to switch between profiles those profiles those profiles could either be
provided by an SMDP+ server you control or by the same provider that provisions
the profiles though the SMDP+ server.

As an alternative to controlling a certified SMDP+ server, an app may also be promoted to
a privileged system app, e.g. by the mobile OS or an OEM, to allow it to 
switch between multiple profiles. 
This is the approach we have taken in the implementation presented here.

The core piece of our implementation is the market where
bandwidth contracts are sold and purchased, discussed next.

\subsection{Blockchain Market}
The key part of the system that allows autonomous
purchasing of bandwidth contracts is a blockchain
digital market where providers can set prices and
UEs can purchase allocations.
The blockchain is implemented as ledger where
bandwidth purchase transactions are recorded
using smart contract processing. We used
the open source Sawtooth Hyperledger\footnote{https://www.hyperledger.org/use/sawtooth} implementation
to implement a custom transaction processor to
verify purchases, atomically execute bandwidth
allocations and offers, and record account balances.
A transaction processor takes a signed payload
and then verifies it against the current state
of the blockchain before adding a verified
transaction to the ledger. We then allow
the UE purchasing a contract to send proof of
purchase to a network provider to get access,
either by directly getting access to AKA parameters
or in the eSIM case by simply enabling the pre-downloaded
eSIM profile in the HSS. All services are implemented
as REST endpoints offering a JSON API.
We also implemented an exchange that allows for
payment gateways to either withdraw or deposit real
currency out of or into the bandwidth ledger.
Each UE and each provider will have a unique account
in the blockchain that in the UE case needs to be initiated
with funds to start executing transactions.

The allocation and verification can also be done in a single step
where the UE will prepare and sign a purchase request transaction
and send that directly to the provider who will forward it to and
execute it on the blockchain, before verifying the transaction and
giving the UE access. This allows the UE to make allocations without
being connected to or having direct access to the blockchain services.

The payloads are encoded as Protocol Buffer bytestreams using
a standard Sawtooth format that allows for batches of transactions
to be encoded and forwarded by third parties. The inner part of the
payload is specific to the transaction processor that we defined.
It can be encoded as json or as a simple comma separated string.
Our custom payload as an action element that defines the intent
of the transaction. It can be {\it allocate}, {\it offer}, {\it deposit},
or {\it withdraw}. Each of these payloads will also have a signer
that requests the action and a target provider that the action
is targeted at. The {\it deposit} and {\it withdraw} actions
can only be performed by trusted exchanges to fund or exchange blockchain
currency to and from other currencies. A UE would typically issue
the allocate actions with a target provider that matches an offer on
the blockchain. The blockchain records that are atomically written
as a result of executed actions provide a cryptographically verified
input payload and a state resulting from executing that payload.
The state in this case provides the most up-to-date record of
the balance of the target account, allocations remaining in an epoch (virtual time).
This state together with the input payload that caused it to be recorded are
available for anyone to verify that has access to the blockchain, i.e.
the network bandwidth providers and payment gateway changes.

The payload is defined in Table~\ref{T:payload}. Note that not all payload
elements are used or required by all actions.
The ledger transaction record is defined in Table~\ref{T:record}, and the
basic processing rules for different actions are defined in Table~\ref{T:actions}. 

\subsection{LTE EPC Integration}
To allow a network provider to sell bandwidth on our blockchain market
they need to interact with the ledger to enable or provision
users on demand and to update pricing and bands of offers. Different
prices may be set on different frequency bands and for different
band width. Each offer configuration has one price. The provider
can also specify how many allocations within an offer can be sold
within an epoch.

Users may purchase allocations independently of the provider and
the provider would then validate proof of a transaction to grant
the user an allocation and access to the network.

We have built two Proof of Concept integrations, and with the EPC/HSS
of srsLTE\footnote{https://github.com/srsLTE/srsLTE}, and one one with the Aricent EPC/HSS\footnote{https://northamerica.altran.com/software/lte-evolved-packet-core}. The srsLTE integration
allows both connected and disconnected allocations over a custom LTE
protocol described below, and it provisions IMSIs and AKA keys on demand.
The Aricent EPC/HSS integration enables and disables pre-provisioned and
pre-downloaded eSIM profiles on demand. 

\subsubsection{srsLTE}
The UE constructs a blockchain transaction by packing a bandwidth
allocation transaction inside a ProtocolBuffer package. It is then
sent over a custom NAS message (see below) to the srs HSS which 
will execute the transaction on the blockchain and if successful 
generate a new IMSI/AKA master key that is sent back to the UE.
The UE will then make a standard NAS Attach call to authenticate.

\subsubsection{Aricent}
In the Aricent integration the users are pre-provisioned in the 
HSS based on allocated eSIM profiles. All IMSIs are set to disabled
before any allocations are made. An integration service also has 
a mapping between the ICCID and IMSIs that are provisioned. When
a phone with an eSIM profile wants to make an allocation they will
like in the srsLTE case construct a ProtocolBuffer blockchain 
transaction locally and send it to the integration service to execute
the transaction together with the ICCID of the profile that
should be enabled. The mapping from ICCID to the correct integration
server representing a network is done via discovery service where
ICCID prefixes may be mapped to the correct provider to request
the allocation from based on a GPS geo-search.
If the transaction executed successfully (price,epoch and provider match
and the signer has enough funds) the IMSI corresponding to the provided ICCID
will be enabled in the Aricent HSS using the Aricent REST API.
The integration service also defines the actual lengths of an epoch.
In our implementation an epoch is by default 5 minutes, which means that
the phone may use the eSIM profiles for 5 minutes before they need to make a new purchase. In this case only connected allocations are supported, similarly to
how eSIMs are provisioned. The connection that is used to make an allocation
may be provided via Wi-Fi or may be an LTE connection that is active.

The integration service we implemented could be seen as a reference implementation of
a network provider integration. To port it to a different network provider
it simply needs to be extended to allow for calls to the EPC/HSS to
enable and disable IMSIs on demand. All the UE and blockchain interactions
can be reused.

\subsection{Android eSIM Agent}
We have implemented the proposed method on the Android platform.
It has been tested with Google Pixel 4 and 4a phones, but should work
with any Android phone that supports eSIM. As alluded to in
Sec.~\ref{sec:esim}, one piece of the implementation runs
as a privileged system app (or carrier app signed with the same
certificate as the eSIM profile) to allow switching between 
eSIM profile providers. This piece has
minimal functionality, whereas the bulk of the implementation
of the agent is implemented as a regular app that could be installed
from the app store. The privileged implementation can only be accessed
by apps that are signed with the same certificate as the privileged app,
mimicking the behavior of carrier apps.

\begin{figure}[htbp]
        \centerline{\includegraphics[scale=0.25]{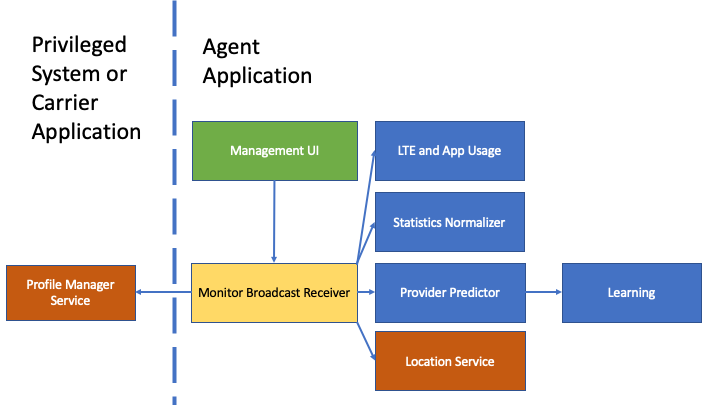}}
        \caption{High-Level Architecture of Android implementation.}
\label{architecture}
\end{figure}

The high-level architecture of the Android app is depicted in Figure~\ref{architecture}.

The {\bf Profile Manager Service}\footnote{Service here refers 
to an Android Service that runs in the background and may be bound to by other services and applications following standard Android semantics} is the privileged system 
component\footnote{e.g., granted privileges using Magisk and systemize, or by matching eSIM certificate} that
is responsible for interacting with the eUICC card, and more specifically instructs it
to activate and deactivate downloaded profiles. It offers an API to external apps and
services that have been signed with the same certificate. 

The component on the Agent Application side of the architecture are all packaged into
the same user-space app and may be deployed from a single apk package, e.g. from
the Google Play Store. The {\bf Management UI} component is the only piece that
runs in the foreground and does not have to be running for the agent to perform
its tasks. It allows users to configured the profiles that the agent should switch between,
e.g. setting data cap and expiration rules, as well as to monitor and visualize 
the learned coverage maps and state transitions, resulting from the learning
algorithm. This component can also pause and resume profile switching, and is
responsible for starting the background task of the agent.

The {\bf Monitor Broadcast Receiver}\footnote{Following standard Android semantics of a broadcast receiver background process} is the entry point for all background tasks.
It is scheduled and wakes up at regular intervals, in our case once a minute regardless whether
the UI part of the app is running. It is responsible for collecting state information and feeding
it to the learning implementation in the form of a reward function.
It calls the {\bf LTE and App Usage} component to collect information about the 
current LTE signal quality and volume of
data (in bytes) transmitted and received with the current LTE provider since the last period (minute),
as well as the currently used (in the foreground) application. It also collects the current
location through a foreground service\footnote{An Android concept that allows services to act like foreground apps to notify users that they perform work in the background, which allows us to collect location
updates as frequently as a foreground app, i.e. once a minute as opposed to 1-2 times per hour
as is the case for standard background apps} called the {\bf Location Service}.
Based on these collected state parameters the {\bf Statistics Normalizer} normalizes the values
to simplify valid comparisons between providers. For example the throughput is normalized based
on which application is running and the signal strength is averaged across location grid cells
computed with the Geohash algorithm\footnote{Splits latitude and longitude coordinates in two planes and assigns bits depending on which side of the demarcation the location belongs, 
and then constructs a string of characters using an alphabet on a byte-by-byte basis. 
We use 5 character strings which roughly corresponds to cells of rectangular cells of 2x2 miles depending on where you are on earth}. 
The former is done to determine whether a spike in
throughput is due to the app or a better provider and the latter is to avoid oscillation
between provider when signal strength fluctuates rapidly, e.g. due to moving at high speed.
Finally the normalized statistics are fed into the reward function defined in Sec.~\ref{sec:reward2},
through the {\bf Provider Predictor} component, which maintains historical state for learning. 
The Provider Predictor then predict based on the learned state which provider
should be picked next. At this point the Monitor Broadcast Receiver can make another
check whether the user is actively using the phone (e.g. screen is on) to determine whether
the switch should be enacted. If it should be enacted a call will be made to the
Profile Manager service to switch. Note, if the switch is slow it will impact the
traffic volume collected in the next period, which will impact the reward, and thus overhead
in switching is baked into the predictions, to introduce natural inertia in switches.

\subsection{LTE PBCH/NAS Extensions}
To prototype allocations in a disconnected state we implemented a new LTE protocol using
a USRP open source implementation called srsLTE. The implementation was tested with
Ettus Research USRP B200 boards\footnote{https://www.ettus.com/all-products/ub200-kit/} both on the UE and eNodeB side. To communicate
bandwidth offers on the blockchain to the UEs in a disconnected state we made use of
the PBCH and a new custom SIB. The SIB broadcast content is listed in Table~\ref{T:sib}.

To purchase an allocation two new NAS messages were defined {\it RequestBandwidthAllocation} and
{\it BandwidthAllocationResponse} shown in Tables~\ref{T:rba} and \ref{T:bar}.

We modified both the srsLTE UE and EPC HSS implementation to send and receive the SIB and NAS messages as well as to install new 
credentials on demand based on successful transactions. The end-to-end transaction had an overhead of about 1s\footnote{srsLTE
was not designed for dynamic AKA credential provisioning so bringing up a new UE stack with a new IMSI configuration had an overhead of about 10s}.

% !TEX root = aim.tex

\section{Experiment}\label{sec:experiment}
We designed an experiment testbed to verify simulation
results with real LTE networks and real phones.
The key difference to the simulations is that the UEs
are not mobile, but app demand and price dynamics as
well as competing agents are reproduced. Furthermore,
utility values are not computed but measured based on
real competing throughput and contention on the
networks with natural variance.
As a result training period impact also becomes more interesting
to study.
Another critical difference between the simulations and experiments
is that the experiments actually purchase bandwidth contracts
on the blockchain market whenever a network selection is made
in each step. The allocation also results in the eSIM being
enabled in the HSS for a limited time (a few minutes) enough
to complete the step transmission.
\subsection{Setup}
We set up two independent LTE networks on different
UL and DL central frequencies, and deployed three
phones with eSIM profiles for both networks. The phones
are stationary and receive perfect signal strength
from both networks, however, one network offers lower bandwidth
than the other. There is no other interference and no other
users of the network than the three phones.

Two Pixel4a phones and one Pixel4 phone made up the UEs in the
experiments. All phones ran the Android 10 OS.

The setup configuration is summarized in Table~\ref{T:netconfig}.
Note, the phones are co-located to the received power and signal
from the two networks are the same for all phones. Likewise
upload and download speeds are identical across all phones
for the two networks.

\begin{table}[htbp]
        \caption{Testbed Network Configuration and Capacity.}
\begin{center}
\begin{tabular}{|l|l|}
\hline
\textbf{\it Network 1} &  \\
\textbf{EARFCN} &  40620 \\
\textbf{Bandwidth (Mhz)} & 10 \\
\textbf{Distance to UEs (ft)} & 4 \\
\textbf{RSRP (dBm)} & -85  \\
\textbf{Max Upload (Mbps)} & 2 \\
\textbf{Max Download (Mbps)} & 32  \\
\hline
\textbf{\it Network 2} &  \\
\textbf{EARFCN} &  39750 \\
\textbf{Bandwidth (Mhz)} & 20 \\
\textbf{Distance to UEs (ft)} & 1\\
\textbf{RSRP (dBm)} &  -72 \\
\textbf{Max Upload (Mbps)} &  9 \\
\textbf{Max Download (Mbps)} & 84  \\
\hline
\end{tabular}
\label{T:netconfig}
\end{center}
\end{table}

\subsection{Experiment Design}
The three phones are connected via USB cables to a central controller
PC in a LAN. The PC also runs iperf3 servers and an ADB bridge to control
the phones. The phones have an iperf3 binary as well as eSIM profiles pre-downloaded
for both networks. The phones also run a carrier app that allows programmatic
switching between the eSIM profiles. 
The phones can discover the networks and their dynamic pricing via a discovery
and bandwidth market we implemented (using blockchains, see implementation section).
The phones simulate app demand by using different iperf3 bitrate limited TCP
transmission or by using unlimited transmission. 
The task of a provider selector is to pick the best provider given
current network prices and app demand. Historical throughput data may be collected
and used to inform the selection in a training phase.

\begin{figure}[htbp]
        \centerline{\includegraphics[scale=0.25]{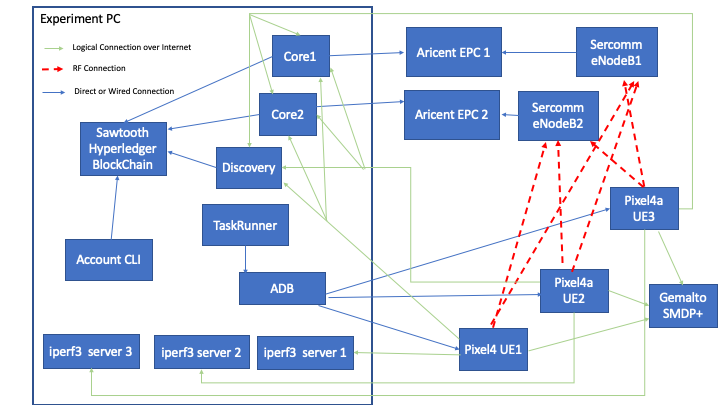}}
        \caption{Architecture of components involved in eSIM testbed.}
\label{testbedarch}
\end{figure}

The architecture of the experiment testbed is depicted in Figure~\ref{testbedarch}.

As with the simulations, app demand and price dynamics are configured similarly.
Both networks alternate their offered price between high (\$4) and low (\$1) 
with 50\% probability for each price.

We ran all base experiments for 30 steps with demand potentially changing for each step.
The first 5 steps were used for training. In each step a 10 second upload is performed,
and the throughput is measured. The throughput is then evaluated against the utility
function of the current application. Like in the simulations we make use of batch and
interactive apps, where interactive apps have a threshold that caps the utility.
Apart from varying the number of UEs that dynamically pick a provider dynamically,
we also vary the application demand to ensure there is an opportunity to {\it pack}
the workloads efficiently across the networks. The demand configuration will be specified
separately for each experiment below.

\subsection{Results}
Before running the provider selection experiments, we first study the contention between UEs with different
traffic types across the networks, we then look at the performance of
selection agents in scenarios where there is no competition between
agents and increasing levels of competition.
\subsubsection{Contention Experiment}
In this experiment we compare the throughput performance 
of uploads and downloads (using iperf3), video streaming and
Web page loading, as well as a workload with a mix of uploads and downloads.
We compare the single UE performance to the performance with competing UEs (1 or 2).

\begin{figure}[htbp]
        \centerline{\includegraphics[scale=0.5]{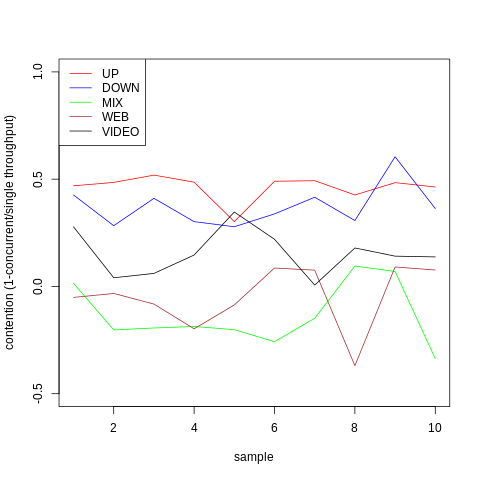}}
        \caption{Contention on a single network with two connected UEs.}
\label{contentionsinglenet2phones}
\end{figure}
\begin{figure}[htbp]
        \centerline{\includegraphics[scale=0.5]{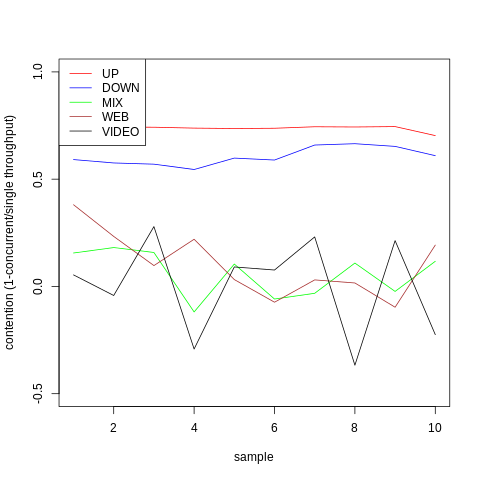}}
        \caption{Contention on a single network with three connected UEs.}
\label{contentionsinglenet3phones}
\end{figure}
\begin{figure}[htbp]
        \centerline{\includegraphics[scale=0.5]{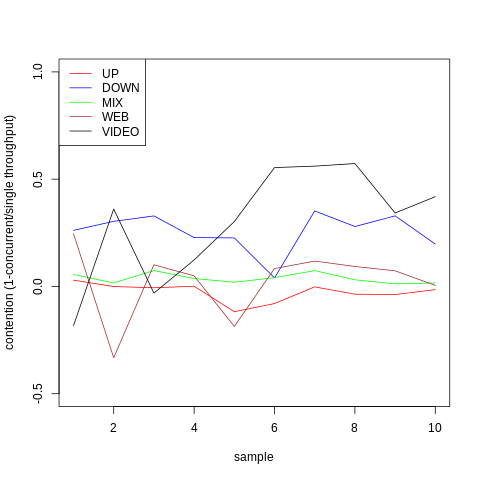}}
        \caption{Contention on two networks with a single connected UE each.}
\label{contention2nets2phones}
\end{figure}

From these experiments we can see that the video and web workloads do not fully stress
the network capacity available and contention is reduced. Download has less contention
than uploads because data are generated on a single PC for download and hence
CPU contention masks network contention. The clearest contention results where
performance degrades with $1/n$ where $n$ is the number of competing UEs is uploads.
Hence, we will utilize uploads in our experiments below.

\subsubsection{Single UE Agent Experiment (1DUT)}
\begin{figure}[htbp]
        \centerline{\includegraphics[scale=0.5]{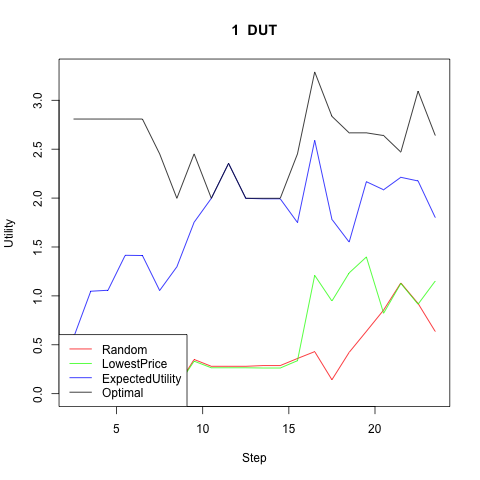}}
        \caption{1 UE running agent, 2 UEs with fixed network and load.}
\label{exp1util}
\end{figure}

In the first experiment we let one UE select an optimal
network to connect to given application demand and
network prices, whereas two UEs are statically connected to
one network each and run a unrestricted TCP upload.
The demand varies randomly between a throughput limited
upload and an unrestricted upload. Figure~\ref{exp1util}
shows the results with 5 training steps and 25 prediction steps.
Both price and demand may change in each step. Price varies between
a premium price and a low price in a bi-modal distribution on
both networks. Utility for the unrestricted demand application
is simply throughput over price, and for the restricted demand
there is a cap on throughput, beyond which utility does not
increase and below which utility has a minimal fixed value.
In this and subsequent experiments only the UEs that select
their network dynamically are included in the aggregate
utility metric. The other UEs simply serve as background load.

The demand configuration for this experiment is shown in Table~\ref{T:demand1}.
\begin{table}[htbp]
        \caption{Demand Configuration: Single UE Agent Experiment (1DUT).}
\begin{center}
\begin{tabular}{|l|l|}
\hline
\textbf{\it UE 1} &  \\
\textbf{Dynamic selection} &  yes \\
\textbf{App 1 Utility Function} & batch \\
\textbf{App 2 Utility Function} & interactive \\
\textbf{App 2 Threshold Demand (Mbps)} & 2 \\
\textbf{Transition Probabilties} &  \\
\textbf{App1$\rightarrow$App1,App1$\rightarrow$App2} & $0.2,0.8$ \\
\textbf{App2$\rightarrow$App1,App2$\rightarrow$App2} & $0.2,0.8$ \\
\hline
\textbf{\it UE 2} &  \\
\textbf{Dynamic selection} &  no \\
\textbf{Network} &  1 \\
\textbf{App Utility Function} & batch \\
\hline
\textbf{\it UE 3} &  \\
\textbf{Dynamic selection} &  no \\
\textbf{Network} &  2 \\
\textbf{App Utility Function} & batch \\
\hline
\end{tabular}
\label{T:demand1}
\end{center}
\end{table}

\subsubsection{Mixed UE Agents Experiment (2DUT)}
\begin{figure}[htbp]
        \centerline{\includegraphics[scale=0.5]{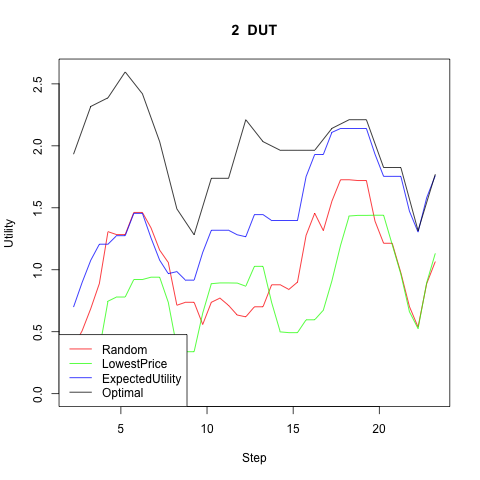}}
        \caption{2 UEs running agent, 1 UE with fixed network and load.}
\label{exp2util}
\end{figure}

In this experiment we let two UEs select network providers in each
step and only one UE is fixed. Figure~\ref{exp2util}
shows the results. 

The demand configuration for this experiment is shown in Table~\ref{T:demand2}.
\begin{table}[htbp]
        \caption{Demand Configuration: Mixed UE Agent Experiment (2DUT).}
\begin{center}
\begin{tabular}{|l|l|}
\hline
\textbf{\it UE 1} &  \\
\textbf{Dynamic selection} &  yes \\
\textbf{App 1 Utility Function} & batch \\
\textbf{App 2 Utility Function} & interactive \\
\textbf{App 2 Threshold Demand (Mbps)} & 2 \\
\textbf{Transition Probabilties} & \\
\textbf{App1$\rightarrow$App1,App1$\rightarrow$App2} & $0.8,0.2$ \\
\textbf{App2$\rightarrow$App1,App2$\rightarrow$App2} & $0.8,0.2$ \\
\hline
\textbf{\it UE 2} &  \\
\textbf{Dynamic selection} &  yes \\
\textbf{App 1 Utility Function} & batch \\
\textbf{App 2 Utility Function} & interactive \\
\textbf{App 2 Threshold Demand (Mbps)} & 2 \\
\textbf{Transition Probabilties} & \\
\textbf{App1$\rightarrow$App1,App1$\rightarrow$App2} & $0.2,0.8$ \\
\textbf{App2$\rightarrow$App1,App2$\rightarrow$App2} & $0.2,0.8$ \\
\hline
\textbf{\it UE 3} &  \\
\textbf{Dynamic selection} &  no \\
\textbf{Network} &  2 \\
\textbf{App Utility Function} & batch \\
\hline
\end{tabular}
\label{T:demand2}
\end{center}
\end{table}

\subsubsection{Only UE Agents Experiment (3DUT)}
\begin{figure}[htbp]
        \centerline{\includegraphics[scale=0.5]{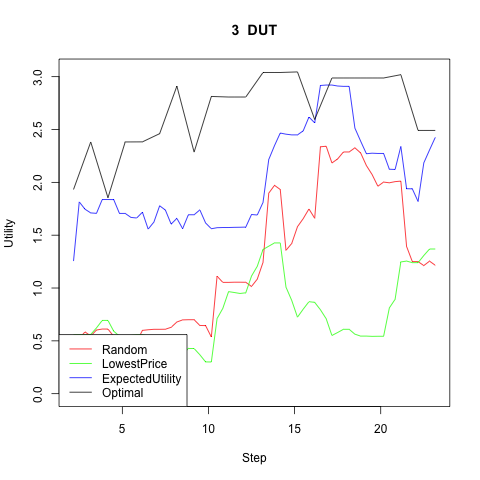}}
        \caption{3 UEs running agent.}
\label{exp3util}
\end{figure}

In this experiment we let all three UEs select network providers in each
step. Figure~\ref{exp3util}
shows the results. 

From these experiment we can conclude that the ExpectedUtility method
accounts for both price and demand to improve the utility or user experience
both in single agent and competing agent scenarios. As expected simply picking
the lowest priced provider does not work well when all agents do the same.

The demand configuration for this experiment is shown in Table~\ref{T:demand3}.
\begin{table}[htbp]
        \caption{Demand Configuration: Only UE Agent Experiment (3DUT).}
\begin{center}
\begin{tabular}{|l|l|}
\hline
\textbf{\it UE 1} &  \\
\textbf{Dynamic selection} &  yes \\
\textbf{App 1 Utility Function} & batch \\
\textbf{App 2 Utility Function} & interactive \\
\textbf{App 2 Threshold Demand (Mbps)} & 2 \\
\textbf{Transition Probabilties} & \\
\textbf{App1$\rightarrow$App1,App1$\rightarrow$App2} & $0.8,0.2$ \\
\textbf{App2$\rightarrow$App1,App2$\rightarrow$App2} & $0.8,0.2$ \\
\hline
\textbf{\it UE 2} &  \\
\textbf{Dynamic selection} &  yes \\
\textbf{App Utility Function} & batch \\
\hline
\textbf{\it UE 3} &  \\
\textbf{Dynamic selection} &  yes \\
\textbf{App 1 Utility Function} & batch \\
\textbf{App 2 Utility Function} & interactive \\
\textbf{App 2 Threshold Demand (Mbps)} & 2 \\
\textbf{Transition Probabilties} & \\
\textbf{App1$\rightarrow$App1,App1$\rightarrow$App2} & $0.2,0.8$ \\
\textbf{App2$\rightarrow$App1,App2$\rightarrow$App2} & $0.2,0.8$ \\
\hline
\end{tabular}
\label{T:demand3}
\end{center}
\end{table}

\subsubsection{Experiment Summary}
Table~\ref{T:experiment_summary} summarizes the results.
Optimal values are computed by taking the optimal
utility (from the best allocation) given demand and price of networks in each step across all
experiment benchmarks. Similarly, selection adjusted values are
computed by taking the optimal utility for the given allocation of a benchmark 
in each step. 

\begin{table*}[htbp]
        \caption{Improvement over random and p-values in t-test.}
\begin{center}
\begin{tabular}{|l|l|l|l|l|l|l|}
\hline
      &  \multicolumn{3}{l|}{\textbf{LowestPrice}} &  \multicolumn{3}{l|}{\textbf{ExpectedUtility}} \\
      &  \textbf{1DUT} & \textbf{2DUT} & \textbf{3DUT} &\textbf{1DUT} & \textbf{2DUT} & \textbf{3DUT} \\
\hline
Random Improvement & 0.28 & -0.19 & -0.33 & 2.8 & 0.41 & 0.71 \\
Selection Adjusted & -0.35 & -0.19 & -0.53 & 0.74 & 0.59 & 0.25 \\
\hline
Optimal Fraction & 0.21 & 0.42 & 0.29 & 0.64 & 0.73 & 0.76 \\
Selection Adjusted & 0.36 & 0.48 & 0.37 & 0.97 & 0.96 & 0.99 \\
\hline
Random t-test p-value & 1 & 0.02 & $3.5 \times 10^{-8}$ & $2 \times 10^{-16}$ & $1.5 \times 10^{-6}$ & $2 \times 10^{-16}$\\
\hline
\end{tabular}
\label{T:experiment_summary}
\end{center}
\end{table*}

\subsubsection{Training Period Impact Experiment}
Given that our approach is based on a learning technique, we 
now study the training period impact on the results.
To limit the variable factors we fix both the prices and the application demand.
As in the 2DUT experiment above we fix one UE on Network 2 and let
2 competing UEs run our agent and learning algorithm with different
number of periods of training. None of the UEs knows the demand of the
other UEs, but their QoE will vary depending on the network selection
of the competing UE. To avoid lockstep allocations the training period
allocations are random on both DUTs.

One UE runs an unlimited batch workload (UE1) and the other an interactive workload 
with a 1Mbps threshold.
The network offering lower bandwidth (Network 1)
charges \$$1$ and the one offering higher bandwidth (Network 2) charges \$$3$ per allocation. 
Given that the prices are fixed, demands are fixed, and
the network capacity is fixed we can easily tell theoretically what the optimal
distribution of UEs on the network is, UE1 on Network 2, and UE2 on Network 1 (see below for more details),
and compute allocation success rate. 

We train the two UEs for different 
training periods and then measure the performance for the first step after the training period.
We repeat the procedure 10 times and measure how many of
these iterations have the optimal distribution of UEs over networks. Note that a random allocator
is expected to provide a perfect allocation 25\% of the time.

Even though we can tell the optimal allocation, the exact QoE delivered will
vary depending on the exact throughput achieved at any given time, and is thus not
fully deterministic, and again, none of the UEs know what demand the other UEs generate or
even which networks they are connected to, and hence each UE needs to 
independently explore and learn the performance on the different networks, as 
well as the behavior of the competing UE and make their
network selection accordingly. 

The experiment configuration is summarized in Table~\ref{T:trainingconfig}.
\begin{table}[htbp]
        \caption{Demand Configuration: Training Experiment.}
\begin{center}
\begin{tabular}{|l|l|}
\hline
\textbf{\it UE 1} &  \\
\textbf{Dynamic selection} &  yes \\
\textbf{App Utility Function} & batch \\
\hline
\textbf{\it UE 2} &  \\
\textbf{Dynamic selection} &  yes \\
\textbf{App Utility Function} & interactive \\
\textbf{App Threshold Demand (Mbps)} & 1 \\
\hline
\textbf{\it UE 3} &  \\
\textbf{Dynamic selection} &  no \\
\textbf{Network} &  2 \\
\textbf{App Utility Function} & interactive \\
\textbf{App Threshold Demand (Mbps)} & 6 \\
\hline
\end{tabular}
\label{T:trainingconfig}
\end{center}
\end{table}

By running all the possible distributions of UEs over the network 
we measured the expected performance for each UE to Network mapping possible.
The results are shown in Table~\ref{T:capacity}.

\begin{table}[htbp]
        \caption{Available Network Capacity (UE1,UE2)}
\begin{center}
\begin{tabular}{|l|l|l|}
\hline
\textbf{UEs} & \textbf{Network 1} & \textbf{Network 2} \\
\hline
1 & $(1.7,1)$ & - \\
2 & $(0.4,1)$ & $(4.5,1)$ \\
3 & - & $(3.8,1)$ \\
\hline
\end{tabular}
\label{T:capacity}
\end{center}
\end{table}

A batch agent (UE1) that samples long enough from this table will determine that
the average throughput from Network 1 is $1.05$Mbps and from Network 2 $4.15$Mbps
with expected utilities $1.05$Mbps/\$ and $1.38$Mbps/\$ and will thus select Network
2. Similarly an interactive workload with threshold 1Mbps will
determine that the average throughput both from Network 1 and 2 is $1$Mbps
and utilities $1$ and $0.33$ and thus will pick Network 1. Hence,
the optimal allocation, as mentioned above, 
would be UE 1 on Network 2 and UE2 on Network 1.

Note, by design if UE1 samples from Network 1 when UE2 is not 
connected to it and from
Network 2 when UE2 is connected to it, it will mistake Network 1 from being better, QoE $1.7$Mbps/\$ versus
$1.27$Mbps/\$. That
means that the agents need to learn not only which network is best but also take competing
UEs into account.

\begin{figure}[htbp]
        \centerline{\includegraphics[scale=0.5]{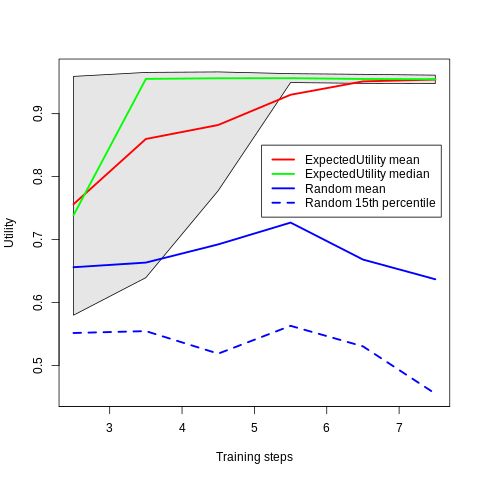}}
        \caption{Training Experiment Utility.}
\label{training}
\end{figure}

Figure~\ref{training} shows the utility results for this experiment,
where the gray area depicts the $70$\% confidence band
of QoE (utility) achieved. The values are smoothed with a two-period moving average. It is easy to see from Table~\ref{T:capacity} that the optimal
utility is $1+1.5=2.5$Mbps/\$, so the utility value in the graph is shown as the fraction of this optimal.

\begin{figure}[htbp]
        \centerline{\includegraphics[scale=0.5]{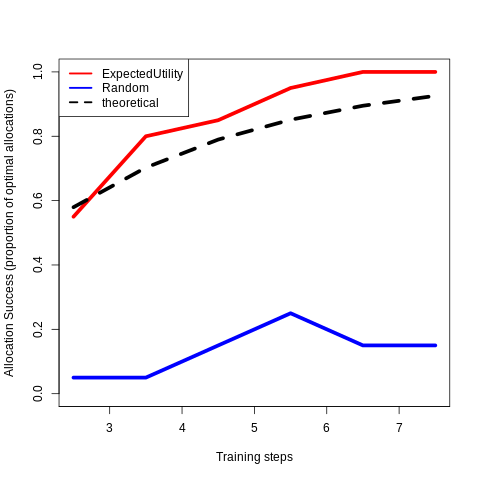}}
        \caption{Allocation Success.}
\label{allocations}
\end{figure}

Figure~\ref{allocations} shows the allocation success rate improvement with increasing
training steps.
With only two training steps the results is comparable to the Random predictor as expected, as the predictor will
make the wrong decision every time UE1 samples from Network 1 when UE2 is not connected to it. It can
happen both in sample 1 and sample 2 with equal probability so about 25\% of the time. These odds are the
same as for the Random predictor, and hence no improvement is shown over Random in this case.

As the number of training samples increases there is less chance of biased sampling from Network 1,
and the expected utility improves. Already with 3 samples there is a drastic improvement to
about $80$\% correct allocations and an average utility fraction of about $85$\% compared to about 
$65$\% with Random. The worst allocations (bottom line of gray area denoting 15th percentile) steadily improves with ExpectedUtility,
and the $70$\% confidence band shrinks and becomes minimal with 90\% correct allocations.
The theoretical allocation success depicted in Figure~\ref{allocations} (again with two period smoothing) 
is computed as $1-\left({\frac{1}{2}}\right)^{.5s}$ where
$s$ denotes the number of samples in the training period. 
We perform better than this theoretical expectation for longer training periods due
to some training periods being used for prediction. We do this to give the UEs a chance to
learn what the preference is of the other UE. 
These results showcase that our method is capable
of learning both network performance and competing user behavior with a small number of samples. 

We should note here that the training samples needed are proportional to the
number of distinct app demands modeled, in this case only one per UE. Hence the training steps needed for
optimal allocations would be longer with more app demands, as you would need to sample each app on each UE
for each combination of app to Network mapping of the other UE. Nevertheless, even if not fully optimal
this mechanism can avoid repeated bad allocations. I.e. if a bad mapping is predicted it will be learned and
not used again. Furthermore, simply assuming either a batch or an interactive demand could also simplify training. 

% !TEX root = aim.tex

\section{Discussion and Next steps}
\label{sec:dualspeedrestless}

\subsection{Pre-paid providers on a dual-SIM device}

We begin with a practical use case that is a special case of the contextual bandit problem.

It is a common usage scenario in many countries for a single user to carry two pre-paid SIM cards (corresponding to two providers) in a single dual-SIM device, and manually enable a SIM (and associated voice and data plan) before making a call or launching an app\footnote{To be precise, the scenario we are describing is that of a ``dual-SIM, single standby'' device as opposed to the newest kinds of devices, which are ``dual-SIM, dual standby,'' where the two SIMs are always enabled and need only be selected by switching amongst them.  However, our analysis applies to both kinds of devices.}.  The reason for this may be that one provider's plan has a higher data cap than another, or has higher data rates than the other, or has more voice minutes included than the other.  Note that this is also the scenario studied in Sec.~\ref{sec:motivation}.

\subsubsection{The provider selection problem}
Clearly, it is tedious for a human user to select and/or enable a SIM before making a call or launching an app.  Let us design a user agent to perform this task automatically on behalf of the human user.  The only action that the user agent is to select the SIM (i.e., provider) to enable next. The two providers have already been paid in advance (through the pre-paid SIM cards), and the selection/enabling of one provider's SIM over the other's does not make any difference to the environment (which here is just the networks of the two providers, and all other users on those two networks). Hence this is a contextual $k$-armed bandit problem with $k=2$.  Thus the analysis of Sec.~\ref{sec:c2ab} applies.  Moreover, as described in Sec.~\ref{sec:state}, the problem can be reformulated with states that can change as a result of the actions of the user agent, and hence direct RL approaches like Q-learning as described in Sec.~\ref{sec:l1rl} are also applicable.

\subsubsection{Reward function(s)}
\label{sec:two_sim_case}
Before defining $R_{t+1}$, we establish the following notation: for any $a \in \{1,\dots,k\}$,
\begin{enumerate}
\item $\mathtt{TotalDataUsed}^{(a)}_t$ are the cumulative data usage on the prepaid plans (as a fraction of their respective data caps); note that $\mathtt{TotalDataUsed}^{(a)}_t = \sum_{t'=1}^t \mathtt{DataUsed}^{(a)}_{t'}$, where $\mathtt{DataUsed}^{(a)}_{t'}$ are the data usage in time step $t'$ on the prepaid plans, as a fraction of their respective data caps;
\item $\mathtt{LifeRemaining}^{(a)}_t$ are the remaining life on the prepaid plans (in units of time steps);
\item $\mathtt{PlanPrice}^{(a)}_t$ are the costs of the prepaid plans (per time step).
\end{enumerate}
We now define $R_{t+1}$ as follows:
\begin{equation}
	R_{t+1} = 
	\frac{\mathtt{QoE}^{(a)}_{t+1}(s)}{\mathtt{PlanPrice}^{(a)}_t} 
	\left(\frac{\mathtt{1 - TotalDataUsed}^{(a)}_t}{\mathtt{LifeRemaining}^{(a)}_t}\right)^\beta
	\label{eq:l1r1}
\end{equation}
for some chosen $\beta \geq 0$, where we explicitly show the dependence of $\mathtt{QoE}^{(a)}_{t+1}$ on the selected provider $a$ and the application $s$ launched by the user, and $\beta \in \{1, 0\}$ depending on whether or not we want to incentivize the use of a provider when its prepaid plan is close to expiry while its data cap has not been reached.  

Alternatively, we may see the existence of the data caps as turning the original contextual $2$-armed bandit problem into a \emph{Contextual Bandit with Knapsack} problem~\cite[Chap.~10]{slivkins2019} defined as follows: at each step $t$, the action $a$ results in the learning algorithm receiving feedback comprising both a reward $R_{t+1} = {\mathtt{QoE}^{(a)}_{t+1}(s)}/{\mathtt{PlanPrice}^{(a)}_t}$ and a (data) consumption vector $(\mathtt{DataUsed}^{(1)}_{t}, \mathtt{DataUsed}^{(2)}_{t})$ with $\mathtt{DataUsed}^{(i)}_{t} = 0$ for $i \neq a$, and the goal is to maximize the (undiscounted) total reward $\sum_{t=1}^{T-1} R_{t+1}$ subject to the budget constraints $\sum_{t=1}^T \mathtt{DataUsed}^{(a)}_{t} \leq 1$, $a=1,2$, where $T$ is the total number of time steps in the lifetime of the plans.  There exist efficient algorithms to solve this convex bandit with knapsack problem that are optimal in a certain sense~\cite{agrawal2016}, but they are more computationally intensive than for the contextual bandit problem.

\subsection{Extensions to the $k$-armed bandit}
We now discuss some extensions and modifications to the basic contextual $k$-armed bandit previously discussed in Secs.~\ref{sec:c2ab} and~\ref{sec:cmab}, and to the RL approach described in Sec.~\ref{sec:duff}.  Although we have not evaluated these methods via simulation or experiment, they are well-known in the literature and are the most promising approaches to study next.

\subsubsection{Contextual UCB bandit}
\label{sec:ucb_bandit}
The upper confidence bound (UCB) modification to the contextual $k$-armed bandit of sec.~\ref{sec:c2ab} changes the right hand side of~\eqref{eq:avs}:
\begin{equation}
	A_t(s) = \arg\max_a \left[Q_t(s, a) + c\sqrt{\frac{\ln t}{N_{t-1}(s, a)}}\right],
	\label{eq:ucb}
\end{equation}
where $c>0$ is a fixed constant~\cite[Sec.~2.7]{sutton2018}.

\subsubsection{Expanding the context space}
Unlike a human user, a user agent residing on the user's device can utilize any attribute or measurement of the network that is available on the device.  The obvious candidates to be used by the agent in choosing an action are the SINRs at the device from the $k$ providers at time $t$, which we denote $\mathtt{SINR}^{(i)}_t$, $i=1,\dots,k$, as these SINRs are not only measured by the device but also strongly influence the QoE of most, if not all, apps.  Note, however, that we can only attempt to predict the QoE at time step $t+1$ based on the SINR measurements at time step $t$.  Moreover, the inherent randomness of wireless channels means that the action $A_t=a$ of selecting provider $a$ does not change the distribution of the SINR $\mathtt{SINR}^{(a)}_{t+1}$, so the SINRs are part of the context but cannot be part of the state.

Let us return to the (non-contextual) $k$-armed bandit problem that, as before, is obtained by treating each context (i.e., launched app) completely separately and decoupled from the other contexts.  Fix the app to be $s \in \{1,2,\dots,n\}$.  This time, however, we shall incorporate the SINRs.  We would expect a larger SINR at time step $t$ from provider $a$ to yield a \emph{preference} on the part of the agent (as it would for a human user) to select the action $A_t(s) = a$.  The so-called Gradient Bandit algorithm expresses this preference through the soft-max function yielding the probability distribution over actions given by~\cite[Sec.~2.8]{sutton2018}
\begin{equation}
	\mathbb{P}\{A_t(s) = a\} = \frac{\exp(H(\mathtt{SINR}^{(a)}_t(s)))}{\sum_{i=1}^k \exp(H(\mathtt{SINR}^{(i)}_t(s)))} = \pi_t(s, a), \text{ say},
	\label{eq:pi1}
\end{equation}
where $H(\cdot)$ is some selected function.
This soft-max has a heuristic physical interpretation for wireless channels in the sense that if noise and interference is constant and fades on the link between the nearest base stations of the different providers and the user device are independent and Rayleigh distributed, then the SINRs from the different providers at the user device are independent and Exponentially distributed.  If $H(\mathtt{SINR}^{(i)}_t(s))$, $i=1,\dots,k$ represent the SINR values in dB, then the above soft-max function $\pi_t(s, a)$ is also the probability that $\mathtt{SINR}^{(a)}_t(s) = \max_{1 \leq i \leq k}\mathtt{SINR}^{(i)}_t(s)$.

The action $A_t(s)$ at time $t$ may be simply drawn from the distribution $\pi_t(s, a)$ or selected as $\arg\max_a \pi_t(s, a)$ or selected as $\arg\max_a \pi_t(s, a)$ with probability $1-\epsilon$ and drawn from the distribution $\pi_t(s, a)$ with probability $\epsilon$ for a chosen small $\epsilon$.

Instead of trying to determine the optimum function $H(\cdot)$ to be used in~\eqref{eq:pi1}, the Gradient Bandit algorithm simply changes the soft-max in~\eqref{eq:pi1} to apply to $H^{(i)}_t$ instead of $H(\mathtt{SINR}^{(i)}_t)$, where $H^{(1)}_t, \dots, H^{(k)}_t$ are called preference functions, and are updated at each time step as follows:
\begin{equation}
	\pi_t(s, a) = \frac{\exp(H^{(a)}_t(s))}{\sum_{i=1}^k \exp(H^{(i)}_t(s))}, \quad a = 1,2,\dots,k,
	\label{eq:pi2}
\end{equation}
where for each $a = 1,2,\dots,k$,
\begin{equation}
	H^{(a)}_{t+1}(s) = 
	H^{(a)}_t(s) + \delta (R_t - \bar{R}_t) [1_{\{a\}}(A_t(s)) - \pi_t(s, a)],	
	\label{eq:gb}
\end{equation}
where $\delta > 0$ is a step size, and $\bar{R}_t$ is either the arithmetic mean, or an exponentially smoothed average, of $R_1,\dots,R_{t-1}$.  We may initialize $H^{(i)}_1(s)$, $i=1,2,\dots,k$ to some function of the SINRs. 

\subsubsection{Extension to the Q-learning direct RL approach}
The Q-learning approach described in Sec.~\ref{sec:duff} is a direct maximization of the expected discounted cumulative reward using RL.  If the SINRs on the links between the user device and the nearest base stations of the provider networks are known, then they can be incorporated into the action selection through
\begin{equation}
	A_t = \arg\max_a \tilde{q}_*(S_t, a) \pi_t(S_t, a),
	\label{eq:L1QLc}
\end{equation}
where $\tilde{q}_*(s, a)$ is as described in Sec.~\ref{sec:l1rl} and $\pi_t(s, a)$ is given by~\eqref{eq:pi1} with the function $H(\cdot)$ selected as $H(x) = \beta \log x$ for some chosen constant $\beta > 0$.  The heuristic reasoning is that the discounted cumulative return $\tilde{q}_*(S_t, a)$ is weighted by the probability $\pi_t(S_t, a)$ that the selected provider turns out to be the one with the highest SINR.

\subsubsection{Modified reward function to account for budget constraints}
The reward function~\eqref{eq:l2r1} for the fixed-price spectrum market of Sec.~\ref{sec:level2} does not account for a budget constraint that may be imposed by the user on the agent's actions.  Of course, a budget constraint may be simply imposed by, say, the payment processing function of the spectrum market, which may refuse to process a bandwidth purchase that leads to the cumulative expenditure by this user exceeding some pre-set limits within a certain pre-set time period.  However, such externally imposed curbs on the agent's decisions will not help the agent learn about these constraints, and may lead to disruptions in the user experience if bandwidth is simply made unavailable to the agent.  Therefore, it is better to incorporate both near-term and longer-term budget constraints into the reward function, so that the agent is able to learn them.  For example, we may modify~\eqref{eq:l2r1} as follows:
\begin{align}
	R_{t+1} &= \frac{\mathtt{QoE}^{(a)}_{t+1}(s)}{\mathtt{PlanPrice}^{(a)}_t} \notag \\
	&\times \max\{0, \text{fraction remaining of} \notag \\ 
	& \qquad \text{extended near-term budget limit}\} \notag \\
	&\times \max\{0, \text{fraction remaining of} \notag \\ 
	& \qquad \text{longer-term budget limit}\}^\beta,
	\label{eq:reward_ext}
\end{align}
where $\beta > 1$ and the ``extended near-term budget limit'' is the near-term budget limit multiplied by the factor $1.1$, say.  This allows the agent to exceed the near-term budget limit (daily or weekly) by up to 10\% so long as the longer-term (monthly) budget constraint is met (which is enforced through the exponent $\beta$).  Note that~\eqref{eq:reward_ext} applies for all agent actions (i.e., bandwidth purchase decisions) that do not breach either the extended near-term limit or the longer-term limit on the budget, for when an agent action would do so, the payment processing function would decline to process the corresponding bandwidth purchase and the reward would therefore be zero.

Alternatively, in the same way as in Sec.~\ref{sec:two_sim_case}, we may formulate the problem with the (strict) long-term budget constraint (now assumed to hold for bandwidth leases from each provider) as a contextual bandit with knapsack problem where the feedback to the algorithm after action $a$ (defined as a set of bandwidth leases from the various providers) in time step $t$ is the reward $R_{t+1} = {\mathtt{QoE}^{(a)}_{t+1}(s)}/{\mathtt{PlanPrice}^{(a)}_t}$ and the consumption vector $\boldmath{c}_t$ whose $i$th entry equals the amount of money spent on bandwidth lease from provider $i$. As discussed in~\cite[Chap.~10]{slivkins2019}, the optimal policy is a randomized one where the actions are drawn from a joint probability distribution over all providers.  The problem can be restated in terms of maximizing the UCB (see Sec.~\ref{sec:ucb_bandit}) of the expected reward with respect to this optimal probability distribution, subject to the relaxed near-term budget constraint of the \emph{lower} confidence bound (LCB) of the expected consumption vector with respect to this probability distribution in each time step not exceeding the budget for that time step (for each provider)~\cite[Alg.~(10.3)]{slivkins2019}. The LCB requirement allows the consumption vector entry corresponding to a given provider in a given time step to exceed the (near-term) budget for that provider for that time step, although the (longer-term) total budget constraint is never relaxed.

\subsubsection{Multi-period selections}
Bearing in mind the overheads and delays each time a user agent switches providers, it makes sense for the agent to stay with a selected provider for more than a single time step.  In this case, the reward function should be the aggregate reward over the individual time steps (with a term to account for the overhead cost of switching), and the action now has to include the selected number of time steps in addition to the selection of the provider.  

\subsection{General provider selection problem as a Restless Bandit}
Recall that in Sec.~\ref{sec:state}, we assumed that the state transitions on the Markov chains $S_t^{(i)}$ for all provider networks $i \in \{1,\dots,k\}$ other than the selected provider $a$ at time $t$ were paused and only the selected provider's Markov chain $S_t^{(a)}$ made a transition to $S_{t+1}^{(a)}$.  This assumption is true for the scenario in Sec.~\ref{sec:state} because the states are just the apps launched by the human user of the device on which the user agent resides.  It even holds for the scenario studied in Sec.~\ref{sec:l2rl} because we assume that all providers hold their prices fixed between two consecutive time steps.  

However, if the providers update their prices asynchronously, which may happen if the providers dynamically update their prices  in response to user agent activity, for example, then we need to relax the restriction that only the selected provider's Markov chain makes a transition and the other Markov chains do not.  The corresponding formulation of the provider selection problem as one of maximizing the expected cumulative discounted reward is called the Restless bandit problem.  The restless bandit has been previously studied in communications in the context of dynamic channel selection~\cite{liu2010, duran2018, wang2018}.

The design and implementation of a user agent for the restless bandit is an open research topic.  Under certain assumptions, the optimal actions for the restless multi-armed bandit problem are obtained from a set of Whittle indices that are similar to the Gittins indices for the (non-restless) multi-armed bandit.  For a system where each provider's state takes only two values, the Whittle indices can be obtained in closed form as shown in~\cite[Thm.~2]{liu2010}.  For richer state spaces, Whittle indices may not even exist, and even if they do, finding them is an open question.  As with Duff's~\cite{duff1995} use of Q-learning to compute Gittins indices, attempts have been made to apply Q-learning to compute Whittle indices~\cite{avrachenkov2020}.  For large state spaces, Q-learning is realized via deep learning models~\cite{wang2018, iclr2021}. However, these deep learning models, because of their higher computational and storage needs, are unsuited for implementation as user agents on mobile devices.  Significant work remains to be done on designing lightweight user agents for the restless multi-armed bandit.

We note in passing that the restless bandit formulation is even applicable to aggregator service providers like GoogleFi if the reward for an individual user is replaced by the total reward to all users served by the aggregator.  This formulation also makes clear the fact that the aggregator cannot consider the value of its actions (provider network selection) for any individual user it serves.

\subsection{The Dual-Speed Restless Bandit problem}
We can model the transition probability matrix in the restless bandit as taking one value when the provider is popular (i.e., selected by more than a minimum number of user agents), and a different value when the provider is unpopular.  This corresponds to a situation where there are more frequent price changes for a popular provider (selected by many user agents) than an unpopular one selected by few user agents~\cite{huberman2008}. The corresponding Bandit problem is called a Dual-Speed Restless Bandit problem\footnote{We assume that the popularity or unpopularity of each provider, as measured say, by the number of user agents selecting that provider's bandwidth offerings exceeding a pre-set threshold, and as represented by a binary state variable, is published and available to all user agents.  Otherwise, the problem becomes one with a partially-observable state, which is considerably more complicated.  Results for a two-armed bandit with partially observed states have been derived in~\cite{fryer2018}.}.  

In fact, we observe that so long as the set of prices charged by the providers is finite, the dual-speed restless bandit problem can describe the general provider selection problem in a spectrum market for bandwidth.  Instead of user agents selecting providers, they actually select a provider's offering, which is represented by a block of bandwidth on a certain frequency band on a specific provider's mobile network.  

As in Sec.~\ref{sec:l2rl}, the entry of the state vector corresponding to the launched app on the selected provider's offering changes, while the corresponding entry in all other providers' offerings does not.  However, unlike in Sec.~\ref{sec:l2rl}, the entries of the state vector corresponding to the prices of \emph{all} offerings by all providers (whether selected by the user agent or not) also have a state transition.  Unlike the definition in Sec.~\ref{sec:state}, the probability transition matrix for these price-state transitions, however, can take not one but two values, $\bm{P}_{\mathrm{pop}}$ and $\bm{P}_{\mathrm{unp}}$ corresponding to ``popular'' and ``unpopular'' providers' offerings respectively, where popularity is defined as being selected by more than a certain threshold number of user agents, say.  As in Sec.~\ref{sec:duff}, we assume that all popular offerings have the same price transition probability matrix $\bm{P}_{\mathrm{pop}}$, and all unpopular offerings have the same price transition probability matrix $\bm{P}_{\mathrm{unp}}$. Suppose there are $n$ possible prices, labeled $\{p_1,\dots,p_n\}$.  A bandwidth offering priced at $p_i$ that drops from the popular to the unpopular category slows down the rate of its price changes by an amount $\epsilon_i < 1$, $i=1,\dots,n$:
\[
	\bm{P}_{\mathrm{unp}} = \mathbf{I}_{n \times n} - \mathrm{diag}(\bm{\epsilon}) + \mathrm{diag}(\bm{\epsilon})\bm{P}_{\mathrm{pop}},
\]
where $\bm{\epsilon} = [\epsilon_1,\dots,\epsilon_n]$.    

Like the restless bandit, the dual-speed restless $k$-armed bandit problem can also be solved using Whittle indices, but the interesting consequence of the dual-speed formulation is that these indices can be computed via the Bertsimas--Ni\~{n}o-Mora algorithm~\cite{bertsimas1996, huberman2008}.  The implementation and evaluation of this approach for a user agent on a mobile device is left for future study.

% !TEX root = aim.tex

\section{Conclusions}
\label{sec:conclusions}
We show that the provider selection problem is naturally formulated as as a restless multi-armed bandit problem.  For a practically relevant and useful scenario, we show that the applicable bandit problem reduces to a contextual multi-armed bandit. We demonstrate via both simulation and experiment that a simple Monte Carlo algorithm for the contextual multi-armed bandit performs well in several scenarios, and also outperforms a direct Reinforcement Learning (Q-learning) approach to maximize the expected cumulative discounted reward.

We have also demonstrated, on a testbed with commercial UEs, the feasibility and benefits of a spectrum bandwidth market allowing
UE agents to self-organize onto the providers offering the best QoE, based on an end-user's
budget, demand and location at any given time, using standard eSIM technology.

Finally, we note that
all these advantages come without compromising the privacy of the user and without central
coordination beyond exchange of digital bandwidth contracts. Furthermore, mobile
network operators do not need to share any information
about their existing users to participate in the market.

\bibliographystyle{plain}
\bibliography{related}

\appendix
\section{Appendix}
\subsection{Implementation Details}\label{sec:implementationdetails}
\begin{table}[htbp]
        \caption{Blockchain Transaction Payload.}
\begin{center}
\begin{tabular}{|l|l|}
\hline
\textbf{provider} & Public key of target of action. \\
\textbf{action} & Allocate,offer,withdraw, or deposit. \\
\textbf{from\_frequency} &  Lower inclusive bound of \\
& frequency in khz being served. \\
\textbf{to\_frequency} &  Upper inclusive bound of \\
&  frequency in khz being served. \\
\textbf{bandwidth} &  Band width within frequency \\
& band in khz being served. \\
\textbf{epoch} &  Monotonically increasing \\
& integer defining the \\
& time period the \\
& allocation is valid. \\
\textbf{price} &  Number of tokens required to \\
& purchase one allocation. \\
\textbf{max\_allocations} &  Max number of allocations \\
& that can be purchased with \\
& the specified bandwidth in \\
& the specified range. \\
\hline
\end{tabular}
\label{T:payload}
\end{center}
\end{table}

\begin{table}[htbp]
        \caption{Blockchain Transaction Processing Semantics for different Actions.}
\begin{center}
\begin{tabular}{|l|l|}
\hline
\textbf{allocate} & Checks that there is a matching \\
& offer in terms of provider, \\
&  epoch and price and that \\
& there are allocations \\
& left to purchase, and that the \\
& signer of the payload has enough \\
& balance in his/her blockchain \\
& account. The price is deducted \\
& from the signer account and \\
& added to provider account on \\
& successful verification. \\ 
\textbf{offer} & Creates a new offer with \\
& a specified price, bandwidth \\
& and freuency band. If the \\
& provider already has an offer \\
& the epoch is incremented, \\
& otherwise it is set to 0.\\
\textbf{withdraw} &  A trusted exchange signer \\
& can request an amount to be \\
& withdrawn from an account.\\
\textbf{deposit} &  A trusted exchange signer \\
& can request an amount to be \\
& deposited into an account. \\
\hline
\end{tabular}
\label{T:actions}
\end{center}
\end{table}

\begin{table}[htbp]
        \caption{Blockchain Transaction Record (State).}
\begin{center}
\begin{tabular}{|l|l|}
\hline
\textbf{provider} & Public key of provider of offer. \\
\textbf{from\_frequency} &  Lower inclusive bound of \\
& frequency in khz being served. \\
\textbf{to\_frequency} &  Upper inclusive bound of \\
&  frequency in khz being served. \\
\textbf{bandwidth} &  Band width within frequency \\
& band in khz being served. \\
\textbf{epoch} &  Monotonically increasing \\
& integer defining the \\
& time period the \\
& allocation is valid. \\
\textbf{price} &  Number of tokens required to \\
& purchase one allocation. \\
\textbf{allocations\_left} &  Number of allocations \\
& left that can be purchased. \\
\textbf{account\_balance} &  Balance of provider on blockchain. \\
\hline
\end{tabular}
\label{T:record}
\end{center}
\end{table}

\begin{table}[htbp]
        \caption{SIB Content.}
\begin{center}
\begin{tabular}{|l|l|}
\hline
\textbf{provider\_public\_key} &  Key used for account \\
 &  in blockchain offer. \\
\textbf{from\_frequency} &  Lower inclusive bound of \\
& frequency in khz being served. \\
\textbf{to\_frequency} &  Upper inclusive bound of \\
&  frequency in khz being served. \\
\textbf{bandwidth} &  Band width within frequency \\
& band in khz being served. \\
\textbf{epoch} &  Monotonically increasing \\
& integer defining the \\
& time period the \\
& allocation is valid. \\
\textbf{price} &  Number of tokens required to \\
& purchase one allocation. \\
\textbf{max\_allocations} &  Max number of allocations \\
& that can be purchased with \\
& the specified bandwidth in \\
& the specified range. \\
\hline
\end{tabular}
\label{T:sib}
\end{center}
\end{table}

\begin{table}[htbp]
        \caption{RequestBandwidthAllocation NAS Message.}
\begin{center}
\begin{tabular}{|l|l|}
\hline
\textbf{batch} &  Base64 encoded UTF-8 of BatchList \\
& message with allocate transaction. \\
\textbf{batch\_id} &  Id of batch (batch header hash) used to \\
& check whether transaction completed. \\
\textbf{timestamp} &  Epoch in seconds since 1970 to protect \\
& against replay attacks. \\
\textbf{nonce} &  Random value to protect against replay \\
& attacks. \\
\textbf{user} &  Public key of user matching signer of \\
& allocate transaction being verified and \\
& used to sign payload. \\
\textbf{signature} &  Secp256k1 signature of preceeding fields \\
& with user key. \\
\hline
\end{tabular}
\label{T:rba}
\end{center}
\end{table}

\begin{table}[htbp]
        \caption{BandwidthAllocationResponse NAS Message.}
\begin{center}
\begin{tabular}{|l|l|}
\hline
\textbf{verified} &  Whether verification succeeded. \\
\textbf{key} &  Master key to be used with AKA on \\
& the UE. Will not be sent if verification\\
& failed. \\
\textbf{imsi} &  IMSI matching MME/HSS PLMN and \\
& associated with the key to be used \\
& with AKA Attach requests on the UE. \\
& Will not be sent if verification failed. \\
\textbf{timestamp} &  Epoch in seconds since 1970 to \\
&protect against replay attacks. \\
\textbf{nonce} &  Random value to protect against replay \\
& attacks. \\
\textbf{signature} &  Secp256k1 signature of preceeding fields \\
& with provider key. \\
\hline
\end{tabular}
\label{T:bar}
\end{center}
\end{table}

%%%%%%%%%%%%%%%%%%%%%%%%%%%%%%%%%%%%%%%%%%%%%%%%%%%%%%%%%%%%%%%%%%%%%%%%%%%%%%%%
\end{document}